\documentclass[aps,twocolumn,showpacs]{revtex4}

\usepackage{epsfig}

\newcommand{\nc}{\newcommand}
\nc{\be}{\begin{equation}}
\nc{\ee}{\end{equation}}
\nc{\bea}{\begin{eqnarray}}
\nc{\eea}{\end{eqnarray}}
\nc{\lsim}{\mbox{\raisebox{-.6ex}{~$\stackrel{<}{\sim}$~}}}
\nc{\gsim}{\mbox{\raisebox{-.6ex}{~$\stackrel{>}{\sim}$~}}}
\nc{\gtwid}{\mathrel{\raise.3ex\hbox{$>$\kern-.75em\lower1ex\hbox{$\sim$}}}}
\nc{\ltwid}{\mathrel{\raise.3ex\hbox{$<$\kern-.75em\lower1ex\hbox{$\sim$}}}}

\begin{document}

\vskip -0.2in

\rightline{CERN-TH/2003-065, HD-THEP-03-13, UFIFT-HEP-03-04}

\vskip 0.2in

\title{Vacuum polarization and photon mass in inflation}

\author{Tomislav Prokopec}
\email{T.Prokopec@thphys.uni-heidelberg.de,
      Tomislav.Prokopec@cern.ch
}
\affiliation{Institut f\"ur Theoretische Physik, Heidelberg
University,
    Philosophenweg 16, D-69120 Heidelberg, Germany}

\affiliation{Theory Division, CERN, CH-1211 Geneva 23, Switzerland}

\author{Richard Woodard}
\email{woodard@phys.ufl.edu}
\affiliation{Department of Physics, University of Florida,
Gainesville, Florida 32611}


\begin{abstract}
We give a pedagogical review of a mechanism through which long wave
length photons can become massive during inflation. Our account
begins with a discussion of the period of exponentially rapid
expansion known as inflation. We next describe how, when the
universe is not expanding, quantum fluctuations in charged particle
fields cause even empty space to behave as a polarizable medium.
This is the routinely observed phenomenon of vacuum polarization.
We show that the quantum fluctuations of low mass, scalar fields
are enormously amplified during inflation. If one of these fields
is charged, the vacuum polarization effect of flat space is
strengthened to the point that long wave length photons acquire
mass. Our result for this mass is shown to agree with a simple
model in which the massive photon electrodynamics of Proca emerges
from applying the Hartree approximation to scalar quantum
electrodynamics during inflation. A huge photon mass is not measured
today because the original phase of inflation ended when the
universe was only a tiny fraction of a second old. However, the
zero-point energy left over from the epoch of large photon mass may
have persisted during the post-inflationary universe as very weak,
but cosmological-scale, magnetic fields. It has been suggested that
these small seed fields were amplified by a dynamo mechanism to
produce the micro-Gauss magnetic fields observed in galaxies and
galactic clusters.
 
\end{abstract}

\pacs{98.80.Cq, 98.80.Hw, 04.62.+v}

\maketitle

\section{Expanding universe and inflation}
\label{Expanding universe and inflation}

The universe is expanding, but with a rate so tiny that it can only
be seen by spectroscopic analysis of stars in distant galaxies.
Suppose the light from such a star contains a distinctive
absorption line measured at the wave length $\lambda$. If the same
line occurs at wave length
$\lambda_E$ on Earth, we say the star's redshift is,
\begin{equation}
z = \frac{\lambda}{\lambda_E} - 1 \,.
\end{equation}
One can also measure the star's flux ${\cal F}$. If we understand the star
enough to know it should emit radiation at luminosity ${\cal L}$, we can 
infer its luminosity distance $d_L$, which is the star's
distance in Euclidean geometry,
\begin{equation}
{\cal F} = \frac{\cal L}{4\pi d_L^2} \quad \Longrightarrow \quad d_L = 
\sqrt{\frac{\cal L}{4 \pi {\cal F}}} \,.
\end{equation}
Astronomers measure the expansion of the universe by plotting $z$ versus
$d_L$ for many stars.

Stars throughout the universe move with respect to their local
environments at typical velocities of about $10^{-3}$ the speed of
light $c$. This motion gives rise to a special relativistic Doppler
shift of $\Delta z \sim \pm 10^{-3}$. If spacetime was not
expanding, this shift would be the only source of nonzero $z$, and
averaging over many stars at the same luminosity distance would give
zero redshift. That is just what happens for stars within our
galaxy. However, the luminosity distances of stars in distant
galaxies are observed to grow approximately linearly with their
redshifts,
\begin{equation}
c^{-1} H_0 d_L = z + \frac12 (1 - q_0) z^2 + O(z^3) \,.
\label{Hplot}
\end{equation}
Equation~(\ref{Hplot}) means that more and more distant objects seem
to recede from us with greater and greater speed. A common analogy is
to the way fixed spots move apart on the surface of a balloon that
is being blown up.

The constant, $H_0 \simeq 2.3 \times 10^{-18}~{\rm Hz}$, is called
the Hubble parameter. (When expressed in units used by
astronomers, $H_0 \simeq 71$\,km/s/Mpc, where 1\,Mpc
corresponds to the distance light traverses in about 3.26 million
years.) Its name honors Edwin Hubble, who established the (nearly)
linear relation~\cite{Hubble:1929} in 1929 based on his
observations, and on earlier work of Slipher and
Wirtz~\cite{Wirtz:1922+1924}. The other constant in Eq.~(\ref{Hplot}) is
known as the {deceleration parameter, $q_0$. Observations of
Type Ia supernovae (whose luminosities can be precisely inferred)
up to the enormous redshift of 1.7 indicate $q_0 
\simeq -0.6$~\cite{Perlmutter-etal:1998,Riess-etal:1998}.

The geometry of spacetime is described by a symmetric tensor field 
$g_{\mu\nu}(x)$ known as the metric. It is used to translate the 
coordinate labels of points $x^{\mu} = (ct,\vec{x})$ into physical 
distances and angles. For example, the square of the distance
between 
$x^{\mu}$ and an infinitesimally close point $x^{\mu} + dx^{\mu}$ is
known as the invariant interval,
\begin{equation}
ds^2 \equiv g_{\mu\nu}(x) dx^{\mu} dx^{\nu} \,.
\end{equation}
Note that we employ the Einstein summation convention in which
repeated indices are summed over $0, 1, 2, 3$.

The transition from nearby stars, whose redshifts are dominated by local 
motions, to more distant stars which obey Eq.~(\ref{Hplot}), is
known as entering the Hubble flow. It is typical in
cosmology to ignore local features and model a simplified
universe that has only the overall expansion effect. Such a
universe does not change in moving between spatial points at the
same time, nor are there any special directions. The first
property is known as homogeneity; the second is 
isotropy.

With a simplifying assumption --- about which more later --- the invariant 
interval of a homogeneous and isotropic universe can be written as
\begin{equation}
ds_{\rm HI}^2 = -c^2 dt^2 + a^2(t) d\vec{x} \cdot d\vec{x} \,.
\label{HI}
\end{equation}
{}From this relation we see that $t$ measures physical time the same
way as in the Minkowski geometry. However, the spatial 3-vector
$\vec{x}$ must be multiplied by $a(t)$ to give physical distances.
For this reason $a(t)$ is known as the scale factor. Its time
variation gives the instantaneous values of the Hubble and
deceleration parameters,
\begin{equation}
H(t) \equiv \frac{\dot{a}}{a}, \qquad q(t) \equiv - 
\frac{a \ddot{a}}{\dot{a}^2} = -1 - \frac{\dot{H}}{H^2} \,.
\end{equation}
The 0 subscripts on $H_0$ and $q_0$ in Eq.~(\ref{Hplot}) and the
subsequent discussion indicate the current values of these
parameters.

Homogeneity and isotropy restrict the stress-energy tensor to only an
energy density $\rho(t)$ and a pressure $p(t)$,
\begin{equation}
T_{00} = -\rho(t) g_{00}, \qquad T_{0i} = 0, \qquad T_{ij} = p(t)
g_{ij},
\end{equation}
where $i$ and $j$ are spatial indices.
In this geometry Einstein's equations take the form,
\begin{eqnarray}
3 H^2 & = & 8 \pi G c^{-2} \rho \,, \label{rhoeqn} \\
-2 \dot{H} -3 H^2 & = & 8 \pi G c^{-2} p \label{peqn} \,,
\end{eqnarray}
where $G$ is Newton's constant. The current energy density is,
\begin{equation}
\rho_0 = \frac{3 c^2 H_0^2}{8 \pi G} \simeq 8.5 \times 10^{-10}~{\rm J/m}^3
\,,
\end{equation}
equivalent to about 5.7 Hydrogen atoms per cubic meter. If we solve
for the instantaneous deceleration parameter,
\begin{equation}
q = \frac12 + \frac{3 p}{2 \rho} \,,
\end{equation}
we find that $p_0 \simeq -0.7
\rho_0$~\cite{Bennett:wmap2003,Spergel:wmap2003}.

By differentiating Eq.~(\ref{rhoeqn}) and then adding $3 H$ times
Eq.~(\ref{rhoeqn}) plus Eq.~(\ref{peqn}), we derive a relation
between the energy density and pressure known as 
stress-energy conservation,
\begin{equation}
\label{12}
\dot{\rho} = - 3 H (\rho + p) \,.
\end{equation}
If we also assume a constant equation of state, $w \equiv 
p(t)/\rho(t)$, Eq.~(\ref{12}) can be used to express the energy
density in terms of the scale factor,
\begin{equation}
\rho(t) = \rho_1 \Bigl(\frac{a(t)}{a_1}\Bigr)^{-3(1+w)} \,.
\label{rhow}
\end{equation}
The substitution of Eq.~(\ref{rhow}) in Eq.~(\ref{rhoeqn}) gives an
equation whose solution is,
\begin{equation}
a(t) = a_1 \Bigl[1 + \frac32 (1 + w) H_1 (t - t_1)\Bigr]^{\frac{2}{3
(1 + w)}} \,. \label{aw}
\end{equation}

The cases of $w = +\frac13$, 0,$ -\frac13$, and $-1$ correspond to
radiation, non-relativistic matter, spatial curvature, and vacuum
energy, respectively. (We do not discuss here scalar field
matter~\cite{RatraPeebles:1987}, usually referred to as
quint\-es\-sence~\cite{CaldwellDaveSteinhardt:1997}, with
nonconstant $w$.) The cosmology for each pure type of
stress-energy can be read off from Eqs.~(\ref{rhow}) and
(\ref{aw}),
\begin{eqnarray}
\mbox{Radiation} & \Longrightarrow & \rho \propto a^{-4},
\quad a(t) 
\propto (H_1 t)^{\frac12} 
,
\nonumber\\ 
\label{raddom} \\
\mbox{Non-Relativistic Matter} & \Longrightarrow & \rho \propto
a^{-3},
\quad a(t) 
\propto (H_1 t)^{\frac23} 
,
\nonumber\\ 
\label{matdom} \\
\mbox{Curvature} & \Longrightarrow & \rho \propto a^{-2}, \quad
a(t) 
\propto H_1 t 
,
\nonumber\\ 
\label{curdom} 
\\
\mbox{Vacuum Energy} & \Longrightarrow & \rho \propto 1, \hskip 2 pc a(t) 
\propto e^{H_1 t} 
.
\nonumber\\ 
\label{dS}
\end{eqnarray}
The actual universe seems to be composed of at least three of the pure types, 
so the scale factor does not have a simple time dependence. However, as long as 
each type is separately conserved, we can use Eq.~(\ref{rhow}) to
conclude that
\begin{equation}
\rho(t) = \frac{\rho_{\rm rad}}{a^4(t)} + \frac{\rho_{\rm mat}}{a^3(t)} +
\frac{\rho_{\rm cur}}{a^2(t)} + \rho_{\rm vac} \,. \label{rhotot}
\end{equation}
As the universe expands, the relative importance of the four types
changes. Whenever a single type predominates, we can infer $a(t)$
from Eq.~(\ref{aw}). This different dependence is one reason it makes
sense to think of an early universe dominated by radiation,
Eq.~(\ref{raddom}), evolving to a universe dominated by
non-relativistic matter, Eq.~(\ref{matdom}). It is also how we can
understand that the current universe seems to be making the
transition to domination by vacuum energy, Eq.~(\ref{dS}).

Under certain conditions there can be significant energy flows
between three of the pure types of stress-energy. For example, as
the early universe cooled, massive particles changed from behaving
like radiation to behaving like non-relativistic matter. This
change would increase $\rho_{\rm mat}$ and decrease 
$\rho_{\rm rad}$ in Eq.~(\ref{rhotot}). The parameter that cannot
change is that of the spatial curvature, $\rho_{\rm cur}$. Strictly
speaking, we should not regard spatial curvature as a type of
stress-energy, but rather as an additional parameter in the
homogeneous and isotropic metric Eq.~(\ref{HI}). We have avoided
this complication because the extra terms it gives in the Einstein
equations (\ref{rhoeqn})--(\ref{peqn}) can be subsumed into the
energy density and pressure, as we have done, and because the
measured value of 
$\rho_{\rm cur}/a_0^2$ is consistent with
zero~\cite{Bennett:wmap2003,Spergel:wmap2003}.

The cosmology in which a radiation dominated universe evolves to matter 
domination is a feature of what is known as the Big Bang scenario. 
Although strongly supported by observation, the composition of $\rho$ at 
the start of radiation domination ($t = t_r$ and $a = a_r$) does not seem 
natural,
\begin{equation}
\rho_{\rm rad} a_r^{-4} \gg \rho_{\rm vac} \gg \rho_{\rm cur} a_r^{-2}
\,. 
\label{hier}
\end{equation}
It might be expected instead that each of the three terms was
comparable, in which case the universe would quickly become
dominated by vacuum energy. There is no accepted explanation for
the first inequality in Eq.~(\ref{hier}) or for the seeming
coincidence that $\rho_{\rm mat} a_0^{-3} \sim \rho_{\rm vac}$.
However, the second inequality in Eq.~(\ref{hier}) finds a natural
explanation in the context of inflation.
 
In 1980 Alan Guth~\cite{Guth:1980} suggested that the Big Bang
scenario was preceded by a period of vacuum energy domination, or
inflation, following which the vacuum energy changed almost
completely into radiation. (Cosmologies that include a period of
vacuum energy domination were independently considered by
Starobinsky~\cite{Starobinsky:1980}, Sato~\cite{Sato:1981}, and by
Kazanas~\cite{Kazanas:1980}.) Suppose that all types of
stress-energy are equally represented at some very early time. We
see from Eq.~(\ref{rhotot}) that the total energy density rapidly
becomes dominated by vacuum energy, following which the scale
factor grows exponentially with a constant Hubble parameter, 
$H_I$. 

The duration of inflation in units of $1/H_I$ is known as the number
of inflationary e-foldings $N_I$. Viable models must have $N_I
\gtwid 50$, and much larger values are common. If $\rho_{\rm
cur}/a^2_I \sim \rho_{\rm vac}$ at the start of inflation,
Eq.~(\ref{rhotot}) shows that the curvature is negligible at the
end,
\begin{equation}
\frac{\rho_{\rm cur}/a_r^2}{\rho_{\rm vac}} \sim 
\Bigl(\frac{a_I}{a_r}\Bigr)^2 = e^{-2 N_I} \ltwid 10^{-44} \,.
\end{equation}
Inflation makes the other types of stress-energy even smaller, but there are 
mechanisms through which vacuum energy can be converted into radiation. This 
process, which we will not discuss, is known as reheating.

Inflation also explains how the large scale universe became so nearly
homogeneous and isotropic. This explanation is crucial because
gravity makes even tiny inhomogeneities grow, and the process has
had 13.7 billion years to operate. It is believed that the galaxies
of today's universe had their origins in quantum fluctuations
of magnitude ${\Delta \rho}/\rho \simeq 10^{-5}$, which occurred
during the last 60 e-foldings of inflation. The imprint of these 
fluctuations in the cosmic microwave background has recently been
imaged with unprecedented accuracy by the WMAP
satellite~\cite{Bennett:wmap2003,Spergel:wmap2003}.

The fact that WMAP did not see the imprint of quantum fluctuations of the 
metric field sets an upper limit of $H_I \ltwid 3.4 \times
10^{38}$\,Hz. No one knows what caused inflation, but a
common assumption is that it occurred at the {grand unified
energy scale $E_{\rm GUT} \simeq 10^{16}\,{
\rm GeV} \simeq 10^6$\,J at which electromagnetic, weak, and strong 
interactions attain equal strength. From Eq.~(\ref{rhoeqn}) this
implies,
\begin{equation}
H_I = \Bigl(\frac{8 \pi G}{3 c^2} \frac{E^4_{\rm GUT}}{(\hbar c)^3}
\Bigr)^{\frac12} \simeq 10^{37}\,{\rm Hz} \,, \label{infHub}
\end{equation}
where we used $\rho_{\rm GUT} = E_{\rm GUT}/V_{\rm GUT}$, 
$V_{\rm GUT} = (\hbar c/E_{\rm GUT})^3$ defines the GUT energy scale volume, 
and $\hbar \simeq 1.05 \times 10^{-34}$\,J\,s is the reduced Planck constant.

\section{Vacuum polarization in flat space}
\label{Vacuum polarization in flat space}

Flat space corresponds to $a(t) = 1$ in Eq.~(\ref{HI}),
\begin{equation}
ds^2_{\rm flat} = \eta_{\mu\nu} dx^{\mu} dx^{\nu} = -c^2 dt^2 + d\vec{x}
\cdot d\vec{x} \,.
\end{equation}
Note that the zero component of a spacetime point $x^{\mu} = (ct,\vec{x})$ 
is $x^0 = ct$, so all components of $\partial_{\mu} \equiv \partial/\partial 
x^{\mu}$ have the dimension of inverse length. Repeated Greek indices are 
summed over $0,1,2,3$ --- for example, $\partial^2 \equiv \partial_{\mu} 
\partial^{\mu}$ --- whereas repeated Latin indices are summed over $1,2,3$ 
--- for example, $\nabla^2 \equiv \partial_i \partial_i$. A dot denotes 
contraction over the appropriate index set, for example, $k \cdot x
\equiv k_{\mu} x^{\mu}$ and $\vec{k} \cdot \vec{x} \equiv k_i x_i$.

To make Lorentz invariance manifest, we employ the Heaviside-Lorentz
system of units in which Maxwell's equations take the form,
\begin{eqnarray}
\vec{\nabla} \cdot \vec{E} = \rho, & & 
\vec{\nabla} \times \vec{B} - \partial_0 \vec{E} = c^{-1} \vec{J}
\,, 
\label{dynam} \\
\vec{\nabla} \cdot \vec{B} = 0, & &
\vec{\nabla} \times \vec{E} + \partial_0 \vec{B} = 0 \,.
\label{nondyn}
\end{eqnarray}
Here $\vec{E}(t,\vec{x})$ and $\vec{B}(t,\vec{x})$ denote the electric and 
magnetic fields, and the charge and current densities are
$\rho(t,\vec{x})$ (for this section only) and $\vec{J}(t,\vec{x})$.
The more familiar MKSA formulation of electrodynamics follows from
the substitutions,
\begin{equation}
\vec{E} \rightarrow \epsilon_0 \vec{E}_{\scriptscriptstyle \rm
MKSA}, \; 
\vec{B} \rightarrow \sqrt{\frac{\epsilon_0}{\mu_0}} \vec{B}_{\scriptscriptstyle
\rm MKSA},\ {\rm and} \; c \rightarrow \frac1{\sqrt{\epsilon_0
\mu_0}} \,,
\end{equation}
where $\epsilon_0$ and $\mu_0$ are, respectively, the electric permittivity
and the magnetic permeability of free space.

It is well known that Eq.~(\ref{nondyn}) can be enforced by
representing the fields using a vector potential $A_{\mu} =
(A_0,A_i)$,
\begin{equation}
E^i = \partial_0 A_i - \partial_i A_0, \qquad B^i =
-\epsilon^{ijk}
\partial_j A_k \,.
\end{equation}
Equations~(\ref{dynam}) combine to the relativistic form,
\begin{equation}
\partial_{\nu} F^{\nu\mu} = c^{-1} J^{\mu} \,, \label{Maxcov}
\end{equation}
using the field strength tensor $F_{\mu\nu} \equiv \partial_{\mu} A_{\nu} - 
\partial_{\nu} A_{\mu}$ and the current 4-vector $J^{\mu} \equiv
(c\rho,\vec{J})$.

Material media such as air and glass consist of an enormous number of atoms 
with negatively charged electrons bound to positively charged nuclei. On 
macroscopic scales such a medium appears neutral and free of currents, but
the application of external fields can distort the bound charges to induce a 
density of atomic electric dipole moments known as the {polarization 
$\vec{P}(t,\vec{x})$, which we illustrate on a gas of polarized atoms 
in figure~\ref{figure-I}. Averaging the actual charge density to remove its 
violent fluctuations on the atomic scale leaves whatever charges are free,
minus the gradient of $\vec{P}$,
\begin{equation}
\langle \rho \rangle = \rho_{\rm free} - \vec{\nabla} \cdot \vec{P}
\,.
\end{equation}
The medium's density of atomic magnetic dipole moments is known as
its magnetization $\vec{M}(t,\vec{x})$. A similar averaging
of the current density gives,
\begin{equation}
\langle \vec{J} \rangle = \vec{J}_{\rm free} + c \partial_0 \vec{P} + c 
\vec{\nabla} \times \vec{M} \,.
\end{equation}
Moving the polarization and magnetization terms to the left-side of
Eq.~(\ref{dynam}) leads to the macroscopic Maxwell equations,
\begin{eqnarray}
\vec{\nabla} \cdot \vec{D} &=& \rho_{\rm free}
\nonumber\\
\vec{\nabla} 
\times \vec{H} - \partial_0 \vec{D} &=& c^{-1} \vec{J}_{\rm free}
\,, 
\label{Maxmed}
\end{eqnarray}
where $\vec{D} \equiv \vec{E} + \vec{P}$ and $\vec{H} \equiv \vec{B}-\vec{M}$.

Linear, isotropic media with no frequency or wave number dependence are 
characterized by,
\begin{equation}
\vec{P} = \chi_e \vec{E}, \qquad \vec{M} = \frac{\chi_m}{1 +
\chi_m}
\vec{B} \,. \label{constchis}
\end{equation}
The dimensionless parameters $\chi_e$ and $\chi_m$ are known as the
electric and magnetic susceptibilities. We would like to
express Eq.~(\ref{Maxmed}) as a single tensor equation like
Eq.~(\ref{Maxcov}). For the case of constant susceptibilities the
result is simple:
\begin{equation}
\partial_{\nu} F^{\nu\mu} + \Pi^{\mu\nu} A_{\nu} = c^{-1} J^{\mu}_{\rm free}
\,. \label{tenseqn}
\end{equation}
where $\Pi^{\mu\nu}$ is the following tensor differential
operator,
\begin{eqnarray}
\Pi^{\mu\nu} &\equiv& \chi_e \Bigl(\eta^{\mu\nu} \partial^2 - 
\partial^{\mu} \partial^{\nu} \Bigr) 
\nonumber\\
 &-& \Bigl(\chi_e + \frac{\chi_m}{1 + \chi_m} \Bigr) \eta^{\mu i} 
\eta^{\nu j} \Bigl(\delta_{ij} \nabla^2 - \partial_i \partial_j
\Bigr) \,.
 \label{Pimunu-constant}
\end{eqnarray}

The polarization tensor encapsulates the medium's effect on
electromagnetic forces and on propagating electromagnetic fields.
It is useful to recall the familiar formulae for the relative
permittivity and permeability and for the index of refraction,
\begin{equation}
\epsilon = 1 + \chi_e, \qquad \mu = 1 + \chi_m, \qquad 
n = (\epsilon \mu)^{\frac12} \,.
\end{equation}
The susceptibility $\chi_e$ in Eq.~(\ref{Pimunu-constant}) 
gives the medium's corrections to the electric
response to a static distribution of charge. Positive $\chi_e$ means that the 
medium's dipoles line up to weaken an applied electric field by a factor of 
$1/\epsilon$. This effect is known as charge screening. The
other term in Eq.~(\ref{Pimunu-constant}) can be understood by
recasting its integrand,
\begin{equation}
\chi_e + \frac{\chi_m}{1 + \chi_m} 
 = \epsilon - \frac1{\mu} 
 = \bigl(n^2-1\bigr)\frac{1}{\mu}
 \equiv \delta n^2\,.
\end{equation}
Together with $\epsilon$, $\delta n^2$ gives the medium's corrections to the 
magnetic response to currents. It also governs the speed $c/n$ at which 
electromagnetic waves propagate, such that, whenever $\delta n^2 \neq 0$ 
($n\neq 1)$, the speed of light differs from that in the (Minkowski)
vacuum.

\begin{figure}[htbp]
\vskip 0.1in
\leftline{\hskip 0.1in\epsfig{file=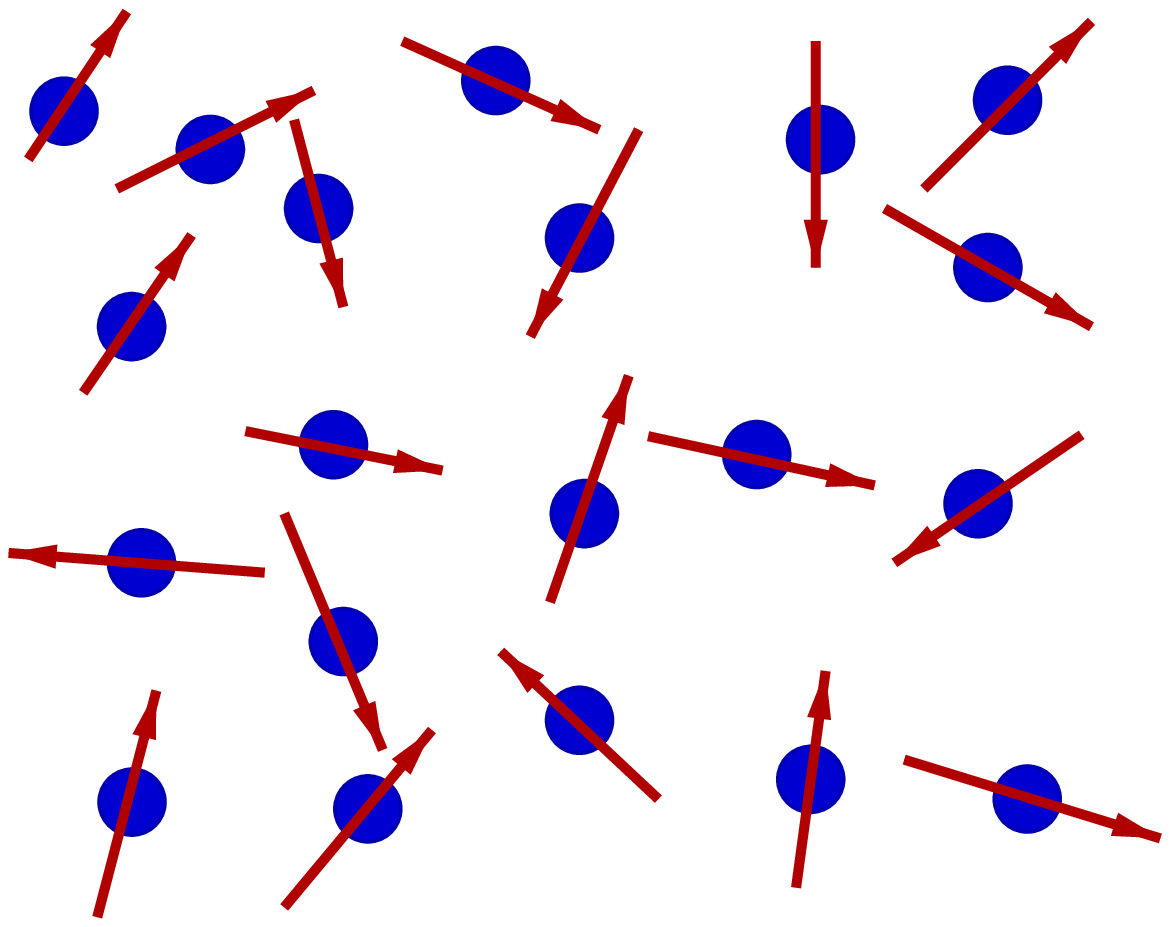,
        width=1.5in,height=1.4in }}
\vskip -1.4in
\rightline{\epsfig{file=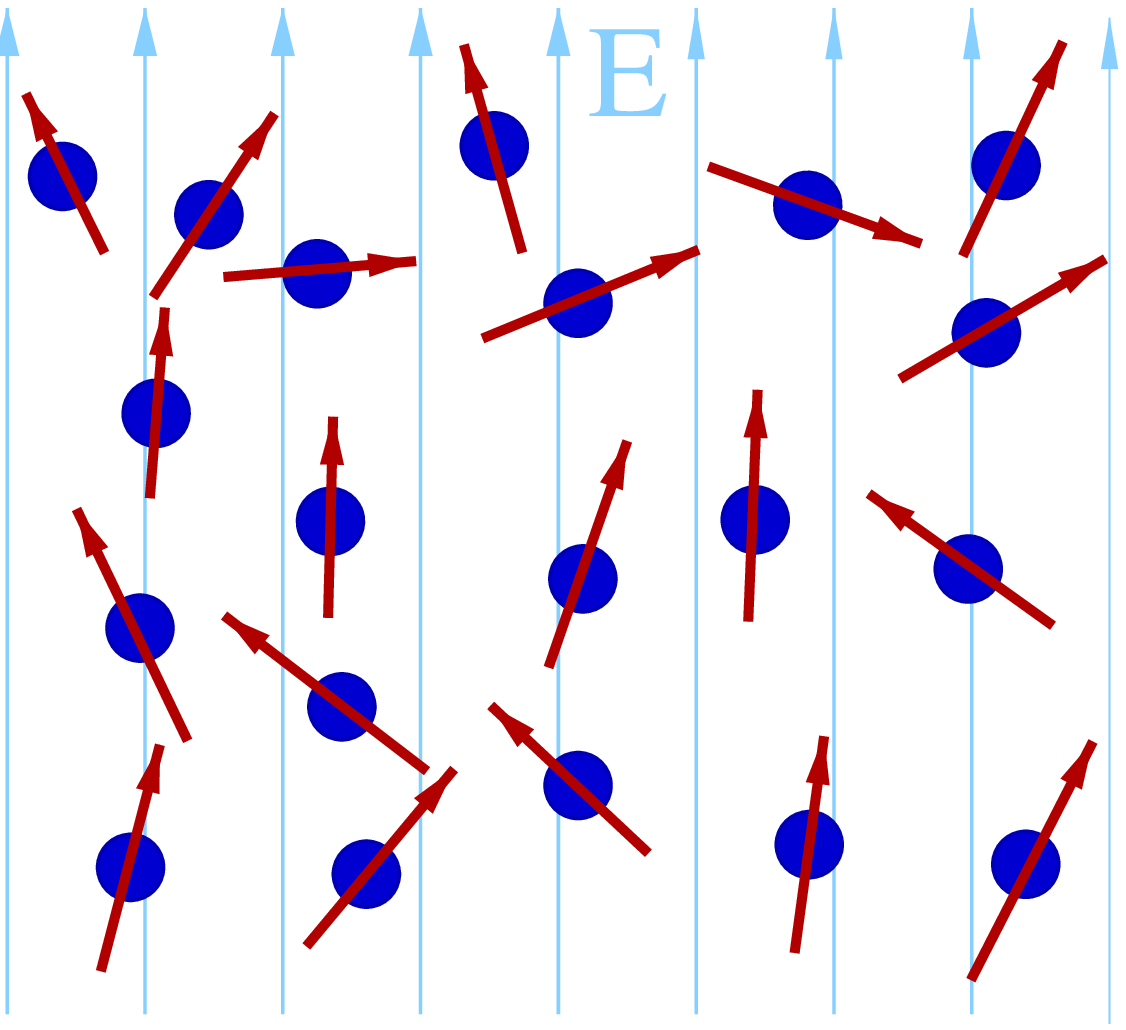, width=1.5in,height=1.4in }
   \hskip 0.1in}
\vskip -0.1in
\caption{A gas of polarized atoms. In the absence of an external
electric field the dipoles orient randomly (FIG.~\ref{figure-I}a). When an
external field $\vec E$ is applied, the dipoles tend to line up
with it (FIG.~\ref{figure-I}b). This alignment produces a net polarization, 
$\vec P = \chi_e \vec E$, which weakens the electric force by $1/(1 +
\chi_e)$. There is a similar effect in vacuum due to oppositely
charged pairs of evanescent, virtual particles. 
\label{figure-I} }
\end{figure}

The susceptibilities of real media typically vary according to the frequency 
and sometimes even the wave number of the external
field~\cite{Jackson:III}. One reason for this variation is that the
bound charges in a medium cannot respond infinitely quickly to
changes in the applied electromagnetic fields, so the polarization
must be attenuated at high frequencies. There also can be resonant
effects when the applied electromagnetic fields perturb systems of 
bound charges near their natural vibrational frequencies. 

Whatever the reason for it, we should understand that frequency
and wave number dependence invalidates the local constitutive
relations (\ref{constchis}) and hence, the tensor equation
(\ref{tenseqn}) we have derived from them. Maxwell's equations in
media (\ref{Maxmed}) are still correct, but the relations between
the polarization and magnetization and the applied fields are only
multiplicative in frequency-wave number space. To calculate
$\vec{P}(t,\vec{x})$ and $\vec{M}(t,\vec{x})$, we first decompose
the electric and magnetic fields into their amplitudes for each wave
4-vector 
$k^{\mu} \equiv (\omega/c,\vec{k})$. This decomposition is
equivalent to taking the Fourier transform, and we denote the
result with a tilde as follows,
\begin{equation}
\widetilde{E}^i(\omega,\vec{k}) \equiv \! \int d^4x e^{i \omega t - i
\vec{k}
\cdot \vec{x}} E^i(t,\vec{x}) \,.
\label{E-fourier}
\end{equation}
To get the polarization, we multiply by the $k^{\mu}$
dependent susceptibility and use the Fourier
inversion theorem,
\begin{equation}
P^i(t,\vec{x}) = \! \int \! \frac{d^4k}{(2 \pi)^4} e^{i k \cdot x}
\chi_e(k) 
\widetilde{E}^i(\omega,\vec{k}) \,.
\end{equation}
The analogous procedure gives the magnetization.

We have just seen that properly accounting for the $k^{\mu}$ dependence of
real media requires us to perform two 4-dimensional integrations:
first over spacetime to Fourier transform the fields, and then over
$k^{\mu}$. This is tedious and unnecessary. By calculating the
Fourier inverse of the susceptibility once and for all,
\begin{equation}
\chi_e(x,x') \equiv \! \int \! \frac{d^4k}{(2\pi)^4} \chi_e(k) e^{i k
\cdot (x - x')} \,,
\label{chi_e:nonlocal}
\end{equation}
we can reduce the process of dealing with different applied electric fields 
to a single spacetime integral,
\begin{equation}
P^i(t,\vec{x}) = \! \int \! d^4x' \chi_e(x,x') E^i(t',\vec{x}')
\,.
\end{equation}
An analogous simplification can be made for the magnetization.

We are now ready to give the generalization of
Eq.~(\ref{tenseqn}) which applies for the case of $k^{\mu}$
dependent media,
\begin{equation}
\eta^{\mu\nu} \eta^{\rho\sigma} \partial_{\rho} F_{\sigma\nu}(x) \!
+ \!\!
\int \! d^4x' [\mbox{}^{\mu} \Pi^{\nu}]\!(x,x') A_{\nu}(x') \! = c^{\!-1} \!
J^{\mu}_{\rm free}(x). 
\label{genMax}
\end{equation}
The polarization bi-tensor has the general form,
\begin{eqnarray}
[\mbox{}^{\mu} \Pi^{\nu}]\!(x,x') & \equiv & - [\eta^{\mu\nu} \eta^{\rho\sigma}
- \eta^{\mu\rho} \eta^{\sigma\nu}] \partial^{\prime}_{\rho} \partial_{\sigma}
\chi_e(x,x') \nonumber \\
&+& \eta^{\mu i} \eta^{\nu j} [\delta_{ij} \partial^{\prime}_k 
\partial_k - \partial^{\prime}_i \partial_j] \delta n^2(x,x') \,,
\quad 
\label{genPi}
\end{eqnarray}
where $\partial_{\mu} = \partial/\partial x^{\mu}$, $\partial_{\mu}^{\prime}
= \partial/\partial x^{\prime\mu}$ and,
\begin{equation}
\delta n^2(x,x') = \!\! \int \!\! \frac{d^4k}{(2\pi)^4} 
\Bigl(\chi_e(k) \!+\! \frac{\chi_m(k)}{\mu(k)} \Bigr) e^{i k \cdot (x - x')} 
\!. \label{deltan2}
\end{equation}
We have actually done a little better than intended. For although
Eqs.~(\ref{chi_e:nonlocal}) and~(\ref{deltan2}) apply only to 
translation invariant and infinite media, 
Eqs.~(\ref{genMax})--(\ref{genPi}) give the response from any linear,
isotropic medium. For example, we could use this formalism to
describe a system in which the local index of refraction varies in
space or even in time. Although we shall not consider spatial
dependence, this natural way of incorporating time dependence will be
quite useful when we consider electrodynamics in an expanding
universe.

The designation of $[^{\mu} \Pi^{\nu}](x,x')$ as a
{\it bi-tensor} derives from general relativity in which the index
$\mu$ transforms according to the vector space at $x^{\mu}$ and the
index $\nu$ according to the vector space at $x^{\prime \mu}$.
Note further that $[^{\mu} \Pi^{\nu}]$ is 
transverse on both indices,
\begin{equation}
\partial_{\mu} [^{\mu} \Pi^{\nu}](x,x') = 0 =
\partial_{\nu}^{\prime} [^{\mu} \Pi^{\nu}](x,x') \,.
\end{equation}

Let us explore the dynamical consequences of an infinite,
translational invariant medium for which
Eqs.~(\ref{chi_e:nonlocal}) and (\ref{deltan2}) pertain. In this
case we may as well Fourier transform (\ref{genMax}) in the Coulomb
gauge, $k_i
\widetilde{A}_i(k) = 0$. The $\mu = 0$ component,
\begin{equation}
-\epsilon(k) \Vert \vec{k} \Vert^2 \widetilde{A}^0(k) = c^{-1}
\widetilde{J}^0(k) \,,
\end{equation}
determines the scalar potential from the charge density. As
claimed, the medium screens the electric forces by a factor of
$1/\epsilon(k)$. The $\mu = i$ equations are more interesting:
\begin{equation}
-\epsilon(k) \Bigl[\frac{\Vert \vec{k} \Vert^2}{n^2(k)} 
 - \frac{\omega^2}{c^2}\Bigr]
\widetilde{A}^i(k) = \frac1{c} \Bigl[ \delta^{ij} - \frac{k^i k^j}{\Vert 
\vec{k} \Vert^2} \Bigr] \widetilde{J}^j(k) \,.
\end{equation}
In addition to the response to a current, the 3-vector potential can also 
support plane waves that obey the following dispersion
relation,
\begin{equation}
\epsilon(k) \Bigl[n^{-2}(k) \Vert \vec{k} \Vert^2 - c^{-2} \omega^2\Bigr] 
= 0 \,. \label{dispersion}
\end{equation}
Einstein's great contribution to quantum theory was the inference (from the
photoelectric effect) that light is quantized in discrete photons of 
energy $E = \hbar \omega$ and 3-momentum $\vec{p} = \hbar \vec{k}$.

When $\epsilon(k)$ is nonsingular, Eq.~(\ref{dispersion}) implies
the usual relation, $\omega = (c/n) \Vert \vec{k} \Vert$. For
this case the energy vanishes as the wave length becomes
infinite. However, suppose the medium obeys,
\begin{equation}
\chi_e(k) = \frac{m_{\gamma}^2 c^2}{\hbar^2 k \! \cdot \! k}, \qquad
n(k) = 1 \,,
 \label{howto}
\end{equation}
where, as we will see in the following, $m_\gamma$ denotes the
photon mass. Although the full significance of the singular
behavior~(\ref{howto}) will become clear later, here we note that
it may arise, for example, in the presence of charged particles
that are constantly created, and which propagate with the speed of
light. From the point of view of the photons, a large number
of these particles may lie on its past light cone, at which $k\cdot k = 0$,
and may thus induce a large (singular) contribution to the susceptibility. 
The substitution of Eq.~(\ref{howto}) in Eq.~(\ref{dispersion})
gives,
\begin{equation}
\epsilon(k) \Bigl[n^{-2}(k) \Vert \vec{k} \Vert^2 - \frac{\omega^2}{c^2}\Bigr] 
= k \! \cdot \! k + \frac{m_{\gamma}^2 c^2}{\hbar^2} = 0 \,.
\end{equation}
Such a photon's energy is that of a massive particle,
\begin{equation}
E = \hbar \omega = \sqrt{\Vert \vec{p} \Vert^2 c^2 + m_{\gamma}^2
c^4} \,.
\end{equation}

We have so far discussed classical media. Quantum field theory predicts that 
particle-antiparticle pairs are continually being created. They live for a 
brief period of time and then annihilate one another. The lifetime of such 
a virtual particle pair is governed by its energy through the 
energy-time uncertainty principle,
\begin{equation}
{\Delta t} {\Delta E} \gtwid \hbar \,, \label{Eprin}
\end{equation}
The meaning of Eq.~(\ref{Eprin}) is that a minimum time ${\Delta
t}$ is needed to resolve the energy with accuracy ${\Delta E}$.
Suppose each partner of a virtual particle pair has energy $E$.
Before they emerged from the vacuum, the energy was zero, whereas
it is $2E$ afterward. This is an example of nonconservation of
energy! However, Eq.~(\ref{Eprin}) says that the violation is not 
detectable in a period shorter than $\hbar/2E$, so virtual
particles can survive roughly that long.

All types of particles experience virtual particle creation with all possible
energies and 3-momenta. One way of understanding the electrostatic force 
is through the exchange of virtual photons. Because normal photons
are massless, they can have arbitrarily small energies and can
therefore survive long enough to mediate the force between distant
charges. However, the lifetimes of massive particles are extremely
short. For example, the minimum energy an electron can have is
that of its rest mass, $m_e c^2 \simeq 8.2 \times 10^{-14}$\,J.
From Eq.~(\ref{Eprin}) we see that an electron-positron pair can 
only live about $\hbar/2 m_e c^2
\simeq 6.4 \times 10^{-22}$\,s. Even moving at nearly the
speed of light (which implies higher energy, and hence shorter
lifetime), a virtual electron would only cover about
$10^{-13}$\,m before annihilating, which is a thousand times
smaller than the scale upon which the discrete electrons and
nuclei are separated in atoms. This explains why macroscopic
experiments do not detect virtual electron-positron pairs. 

Charged virtual particles behave much like the bound charges of
atoms in a polarizable medium. When no external field is present,
the positive partner of a virtual pair emerges as often in one
direction as any other. However, the application of an electric
field makes it preferable for the positive partner to emerge in the
direction of the field, while the negative partner emerges in the
opposite direction. In this way even empty space can acquire a
polarization. The effect is known as vacuum polarization, and
it is described with the same Eqs.~(\ref{genMax})--(\ref{genPi})
which we introduced to quantify the polarization of material media.

Although all charged virtual particles contribute to vacuum
polarization, the largest effect comes from the lightest particles
because they live the longest. Electrons and positrons are about 200
times lighter than the next lightest charged particles, muons, so
almost all vacuum polarization comes from them. By making use of
rather sophisticated techniques of quantum electrodynamics, one can
show that, at lowest order in $\alpha$, they induce the following
electric susceptibility~\cite{PeskinSchroeder:1995},
\begin{eqnarray}
\chi_e(k,\Lambda) 
&=& 8\pi \alpha \! \int_0^1 dx x (1-x)\!\int^\Lambda \!
\frac{d^3p}{(2\pi)^3} \nonumber\\
 &\times& \Bigl[\|\,\vec p\,\|^2 
+ \Bigl(\frac{m_ec}{\hbar}\Bigr)^2 + x(1\!-\!x)k\cdot
k\Bigr]^{-\frac32}
,\quad
\label{QED4a}
\end{eqnarray}
where $\alpha \equiv e^2/4 \pi \hbar c \simeq 1/137$ is the 
fine structure constant. The integral over $\vec{p}$ represents the 
contribution from an electromagnetic field of wave vector $\vec{k}$ exciting
an electron with wave vector $\vec{p} + x \vec{k}$ and a positron with 
wave vector $-\vec{p} + (1-x) \vec{k}$. 

The process described by Eq.~(\ref{QED4a}) conserves 3-momentum
($= \hbar \times {\rm wave\ vector}$), but it need not conserve
energy,
\begin{eqnarray}
\hbar c k^0 &\longrightarrow& \sqrt{m_e^2 c^4 + \hbar^2 c^2 \Vert 
\vec{p} + x \vec{k} \Vert^2} 
\nonumber\\
 &&+\; \sqrt{m_e^2 c^4 + \hbar^2 c^2 \Vert \vec{p} - (1-x) 
\vec{k} \Vert^2}.
\end{eqnarray}
The energy of the electron-positron pair can be very much larger than the
energy of the initial and final photons, which means the pair can
exist only a short time. As one might expect, pairs with very high
$\Vert \vec{p} \Vert$ do not contribute much susceptibility
because they exist too briefly to polarize much. However, there
are so many states with large $\Vert \vec{p} 
\Vert$ that their net contribution diverges. That is why we have
cut the integral off in Eq.~(\ref{QED4a}) for $\|\vec p \|\leq
\Lambda$. If we integrate over the momenta, and expand in
powers of $1/\Lambda$, we obtain, 
\begin{eqnarray}
\chi_e(k,\Lambda) &=& \frac{4\alpha}{\pi} \!\int_0^1 dx x (1-x)
\Bigl[\ln\Big(\frac{2\hbar\Lambda}{m_e c}\Big) - 1 
\nonumber\\
 &-& \frac12 \ln\Bigl(1+x(1-x)\frac{\hbar^2k\cdot k}{m_e^2c^2}\Bigr) 
+ O(1/\Lambda)\Bigr].
\qquad
\label{QED4b}
\end{eqnarray}
We see that the susceptibility diverges logarithmically in the limit
that
$\Lambda\rightarrow \infty$. This is a classic example of an 
ultraviolet divergence in quantum field theory. (The adjective
ultraviolet is used because the problem originates from high
energy --- or ultraviolet --- pairs.)

The infinite susceptibility is not directly observable because it
is a constant, independent of the wave 4-vector $k^{\mu}$ of the
applied electrodynamic field. All observable quantities depend as
well upon the equally constant, bare charges of particles whose
motions reveal the field. For example, consider the force between
two particles of charge $e_{\rm bare}$ that are held a fixed
distance $r$ from one another. Because the system is constant in
time, we have $k^0 = 0$. Because the physical dimension of the 
system is $r$, the maximum response is for wave vectors of about
$\Vert 
\vec{k} \Vert \sim 2\pi/r$. As the separation becomes larger, the
significant wave vectors become ever closer to zero. For very
large separations the force is therefore $1/r^2$ times the
quantity, $e^2_{\rm bare}/4\pi [1 + 
\chi_e(0,\Lambda)]$. The case of large separations (on the scale of
$10^{-13 }$\,m) is one that we can access quite easily, and most
everyone who takes introductory physics performs such a
measurement. Because the result is finite, it follows that the
ratio $e^2_{\rm bare}/[1
+ \chi_e(0,\Lambda)]$ must be a finite number that we call $e^2$,
the square of the measured charge. In other words, the
divergence in the unobservable quantity $e^2_{\rm bare}$ cancels
the divergence in the equally unobservable quantity
$\chi_e(0,\Lambda)$ so that their ratio agrees with what we
measure. To leading order in $\alpha$, the finite $k^{\mu}$
dependent susceptibility, which remains when measured charges are
used, is,
\vskip 0.3in
\begin{eqnarray}
\chi_e(k) &=& \lim_{\Lambda \rightarrow \infty} \Bigl\{ \chi_e(k,
\Lambda) - \chi_e(0,\Lambda)\Bigr\} \\
& =& -\frac{2 \alpha}{\pi} \! \int_0^1 \! dx \, x (1-x) 
\ln\Bigl[1 + x (1-x) \frac{\hbar^2 k \! \cdot \! k}{m_e^2 c^2}
\Bigr].\quad
\nonumber\\
\label{QED4c}
\end{eqnarray}
This discussion illustrates how the process of renormalization works
in quantum field theory. We mention it only to explain why the
finite remainder Eq.~(\ref{QED4c}) can make electromagnetic forces
stronger at short distances. Recall that the least energetic
electron-positron pairs can only survive long enough to travel
about $10^{-13}$\,m. More energetic virtual pairs are limited to
even shorter distances. This means that charged particles separated
by more than about $10^{-13}$\,m feel the polarizations contributed
by virtual pairs of all 3-momenta. But at separations of less than
$10^{-13}$\,m, the lower energy virtual pairs leave the electric
field between the two charges before fully polarizing. The net
effect is less charge screening than at large distances, and hence
a relative enhancement of the electromagnetic force at short
distances.

This effect is known as running of the force law,
and it is seen routinely in precision measurements.
In high energy accelerators such as LEP at CERN and SLC at Stanford
University, charged particles have been brought as close as 
$\sim 10^{-18}$\,m. 
The substitution of $k \! \cdot \! k = (2\pi/10^{-18}$\,m$)^2$
in Eq.~(\ref{QED4c}) gives $\chi_e \simeq -0.023$, or a 2.3\%
enhancement of the electromagnetic force.

If the electron mass had been zero, we would see the
electromagnetic force law run even at macroscopic distances. In
that case the renormalized $e^2$ would be $4\pi R^2$ times the
measured force at some $R$, and this length would enter the
formula for $\chi_e(k)$,
\begin{equation}
\chi_e(k)\Bigl\vert_{m_e = 0} \longrightarrow -\frac{\alpha}{3 \pi} \ln(\mu^2 
k \! \cdot \! k), \quad {\rm where} \quad \mu = \frac{R}{2\pi}.
\label{m=0}
\end{equation}
For $r > R$ we would measure the force to be smaller than $e^2/4
\pi r^2$, whereas it would be greater for $r < R$. The experiment
could be done using the apparatus depicted in
Fig.~\ref{figure-II}~\cite{WilliamsFallerHill:1971,GoldhaberNieto:1971}.

\begin{figure}[htbp]
\epsfig{file=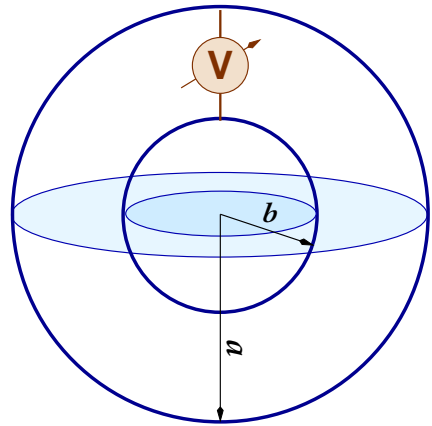, height=2.2in,width=2.2in}
\vskip -0.1in
\caption{A sketch of the Maxwell-Cavendish experiment. Two metal
concentric spheres of radii $a$ and $b$ are first grounded, then the
outer sphere is raised to a high potential. If Cou\-lomb's law were
violated, the voltmeter $V$ would show a non-zero voltage. 
\label{figure-II} }
\end{figure}

We have so far discussed only the term $\chi_e(x,x')$. It turns out that
$\delta n^2(x,x')$ is zero for all relativistic quantum field theories in flat
space. This must be so because the tensor coefficient of $\delta n^2(x,x')$ 
in Eq.~(\ref{genPi}) breaks the Lorentz symmetry between space
and time. One consequence is that the index of refraction is
one, so vacuum polarization does not modify the speed of light.
Because the invariant element of an expanding universe
Eq.~(\ref{HI}) also distinguishes between space and time, we 
might expect that
$\delta n^2(x,x') \neq 0$ when the Hubble parameter is nonzero,
and we will see that this is the case.

Because Eq.~(\ref{QED4c}) has no pole at $k \! \cdot \! k = 0$,
the vacuum polarization from quantum electrodynamics preserves
the photon's zero mass. This turns out to be a
slightly mass and dimension-dependent statement. In 1962 Julian
Schwinger showed that zero mass electrons in one spatial dimension
would induce the following electric
susceptibility~\cite{Schwinger:1962},
\begin{equation}
\chi_e(k) = \frac{4 \alpha}{k \! \cdot \! k} \,. \label{QED2}
\end{equation}
(In two spacetime dimensions $e^2$ has the dimension of ${\rm energy/length}$,
which means that $\alpha$ has the dimension of ${\rm length}^{-2}$.) 
A comparison with Eq.~(\ref{howto}) implies a photon mass of
$m_{\gamma} = 2 
\sqrt{\alpha} \hbar/c$. We will see later that the expansion of
spacetime can also induce a nonzero photon mass.

Electrons and positrons, both of which are spin 1/2 particles,
are not the only kinds of charged particles. To obtain
the susceptibility contributed by other kinds of spin $\frac12$
particles, we simply replace $m_e$ in Eq.~(\ref{QED4c}) by the
appropriate mass. Charged particles with spin zero --- which are
known as scalars --- also entail replacing the
factor of $x (1-x)$ which multiplies the logarithm in
Eq.~(\ref{QED4c}) by $(1-2x)^2/8$. The susceptibility of a zero
mass scalar would be $\frac14$ times that of Eq.~(\ref{m=0}). It
is actually simpler to express the polarization of massless
particles in position space by performing the Fourier transform
of the electric susceptibility, as indicated 
in Eq.~(\ref{chi_e:nonlocal}). The result for a massless, charged
scalar 
is~\cite{ProkopecTornkvistWoodard:2002,ProkopecTornkvistWoodard:2002b},
\begin{equation}
\chi_e(x,x') = -\frac{\alpha}{96 \pi^2} \partial^4 \Bigl\{\! 
\theta(\Delta t) \theta(\Delta \tau^2) [1 - \ln(\nu^2 {\Delta
\tau}^2) ] \!
\Bigr\},
\label{SQED}
\end{equation}
where $\Delta t \equiv t - t'$, $\Delta \tau^2 \equiv (t 
- t')^2 - c^{-2} \Vert \vec{x} - \vec{x}' \Vert^2$ and $\nu 
= c/R$ is the frequency scale of renormalization. Note that 
Eq.~(\ref{SQED}) is zero whenever the point $x^{\prime \mu}$ lies
outside the past light-cone of $x^{\mu}$. This feature, which is
a fundamental requirement on any $\chi_e(x,x')$ and $\delta
n^2(x,x')$, is known as causality.

\section{Virtual particles with expansion}
\label{Virtual particles with expansion}

Leonard Parker was the first to give a quantitative assessment of how the 
universe's expansion can affect virtual
particles~\cite{Parker:1969}. The mechanism is that the partners of
a virtual pair must cover more distance getting back together than
they did moving apart. This causes them to stay apart longer.
Under certain conditions they can become trapped in the Hubble 
flow and become pulled apart, leading to physical particle creation.
The purpose of this section is to explain why the effect is
strongest during inflation and for massless scalars which possess
the special property of {\it minimal coupling} to gravity, about
which more later.

First consider how the energy-time uncertainty principle
Eq.~(\ref{Eprin}) generalizes to the homogeneous and isotropic
geometry in Eq.~(\ref{HI}). Just like photons, a general quantum
mechanical particle is characterized by its wave vector
$\vec{k}$, which points in the particle's direction of
propagation and has magnitude $2\pi$ divided by the particle's
wave length. Now recall from Eq.~(\ref{HI}) that the physical
length between two fixed spatial points is 
$a(t)$ times their coordinate separation. It follows that the physical wave 
vector is $\vec{k}_{\rm ph} = \vec{k}/a(t)$. The 3-momentum of a quantum 
mechanical particle is $\hbar$ times its physical wave vector. Hence the 
energy of a particle with mass $m$ and coordinate wave vector $\vec{k}$ is,
\begin{equation}
E(t,\vec{k}) = \sqrt{m^2 c^4 + \hbar^2 c^2 \Vert \vec{k} \Vert^2/a^2(t)}
\,.
\label{Ewitha}
\end{equation}
This changes with $t$ so the energy-time uncertainty principle says we 
cannot detect a violation of energy conservation at time $t + {\Delta t}$
from a pair of such particles created at $t$, provided that
\begin{equation}
\int_t^{t + {\Delta t}} \!\! dt' \, 2 E(t',\vec{k}) \ltwid
\hbar \,. \label{lifet}
\end{equation}

We see from Eq.~(\ref{Ewitha}) that the growth of $a(t)$ always
reduces the energy relative to the constant scale factor. From
Eq.~(\ref{lifet}) we see that the growth of $a(t)$ always
increases the time a virtual pair can survive. For a given
$\vec{k}$ and time dependence $a(t)$, the rate at which
$E(t,\vec{k})$ falls increases as the mass decreases. Hence,
massless virtual particles experience the largest increase in
their lifetimes.

To understand why inflation maximizes the effect, consider the form
of Eq.~(\ref{lifet}) for a massless particle,
\begin{equation}
2 \hbar c \Vert \vec{k} \Vert \!\int_t^{t + {\Delta t}}
\!\! dt'
\frac1{a(t')} \ltwid \hbar \,. \label{expbnd}
\end{equation}
For the radiation dominated scale factor Eq.~(\ref{raddom}) the
integral grows like $(\Delta t)^{\frac12}$; for matter
domination, Eq.~(\ref{matdom}), its growth is like $(\Delta
t)^{\frac13}$; and the growth is logarithmic for curvature 
domination Eq.~(\ref{curdom}). In each of these cases the
inequality is eventually violated as $\Delta t$ grows. However,
for the inflationary scale factor in Eq.~(\ref{dS}), the integral
approaches a constant as $\Delta t$ becomes infinite. This means
that a long enough wave length pair need never recombine. Models
of inflation are typically well approximated by $a(t) \propto
e^{H_I t}$, for which the bound Eq.~(\ref{expbnd}) takes the
form,
\begin{equation}
2 \frac{c \Vert \vec{k} \Vert}{a(t)} \Bigl[1 - e^{- H_I \Delta t}
\Bigr] 
\ltwid H_I \,.
\end{equation}
Therefore massless particles of coordinate wave vector $\vec{k}$
are created during inflation whenever a virtual pair emerges with
$c \Vert \vec{k}_{\rm ph} \Vert \sim H_I$. This condition on the
wave number is known as first horizon crossing.

\begin{figure}[htbp]
\centerline{\epsfig{file=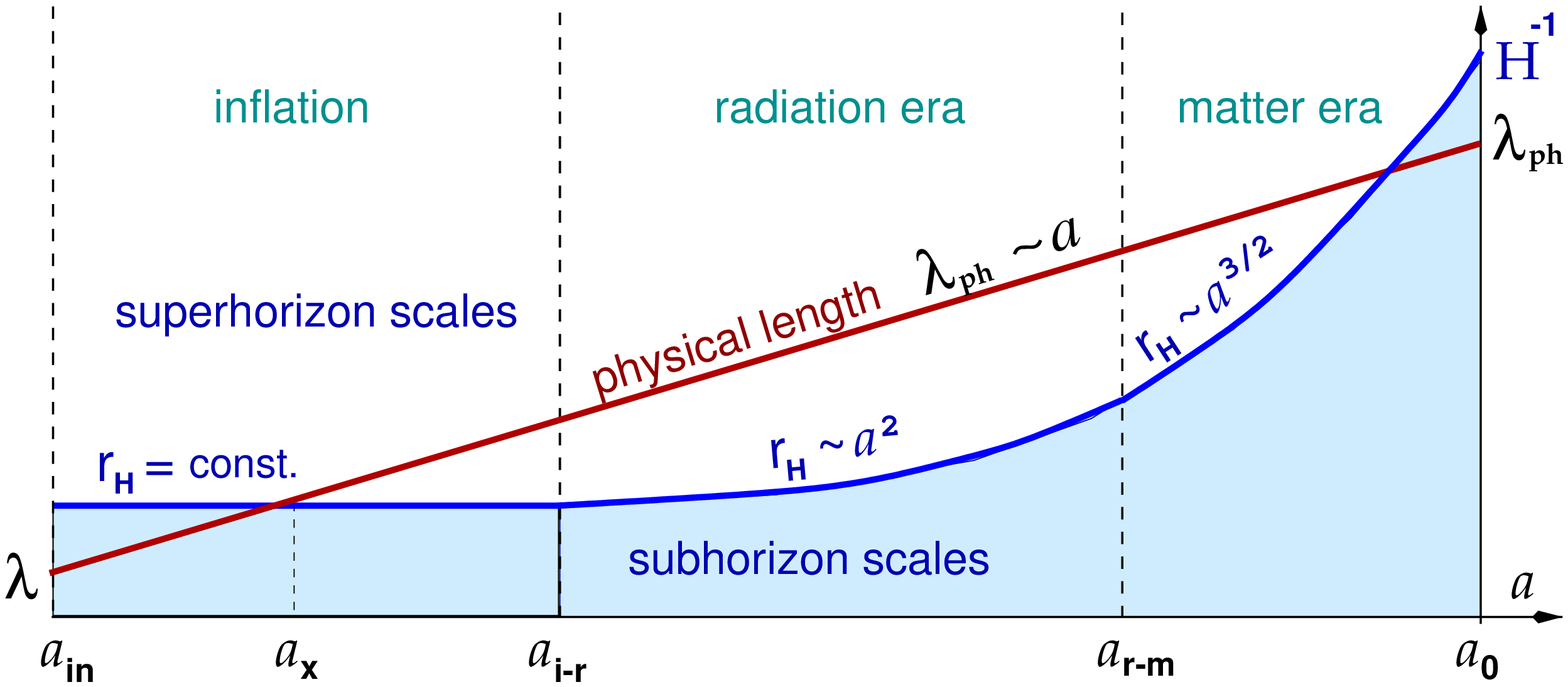,
width=3.4in,height=1.6in}}
\vskip -0.1in
\caption{Evolution of a quantum particle's physical wave length $\lambda_{\rm 
ph} = a(t) \lambda$ as the universe expands. Wave lengths which are
now of cosmological size were originally minuscule. First horizon
crossing occurs (at $a_x$) when $\lambda_{\rm ph}$ becomes
comparable to the inflationary Hubble radius $\sim c/H_I$. 
At first crossing massless, minimally coupled scalars and
gravitons of that wave length are ripped out of the vacuum by the
expansion of spacetime. These particles ride the subsequent
evolution of the universe relatively undisturbed until the second
horizon crossing at
$\lambda_{\rm ph} 
\sim c/H(t)$. Then the particles manifest themselves as
cosmological-scale correlations which could not have formed
causally after inflation. 
\label{figure-III}}
\end{figure}

As indiated in Fig.~\ref{figure-III}, after the first horizon
crossing particles evolve on superhorizon scales during
inflation and subsequent eras, to cross an horizon again in the
radiation or matter era. 
For later use we shall now show how to relate the scale
factor at the second horizon crossing to the coordinate wave
vector $\Vert \vec{k} \Vert$. To this end, it is convenient to
use the redshift, $z \equiv a_0/a(t) - 1$, rather than time to
label events. This choice implies $a = a_0/(1 + z)$. For 
simplicity, we assume perfect matter domination from
matter-radiation equality ($z_{\rm eq} \simeq 3200$) to the present
($z_0 = 0$). From Eq.~(\ref{matdom}) of
Sec.~\ref{Expanding universe and inflation}, we can solve for
the time in terms of the redshift and express the instantaneous
Hubble parameter
$H(t) = 2/3t$ in terms of $z$. If we multiply by the scale factor,
we find
\begin{equation} 
a H = a_0 H_0 \sqrt{1+ z}. \qquad \mbox{(matter domination)} 
\end{equation}
For $z > z_{\rm eq}$ we assume perfect radiation domination.
By following the same procedure with Eq.~(\ref{raddom}), we find,
\begin{equation}
a H = a_0 H_0 \frac{1+ z}{\sqrt{1 + z_{\rm eq}}}. \qquad 
\mbox{(radiation domination)} \label{2nd}
\end{equation}
Suppose galaxies form at $z = 10$. A physical wave length of
10\,kpc would have experienced second horizon crossing
during the radiation epoch at about $z_{2x} \simeq 1.5 \times 10^7$.
Therefore the scales relevant to galaxy formation crossed during
the radiation era, and we conclude that
\begin{equation}
\frac{a_0 H_0 z_{2x}}{\sqrt{z_{\rm eq}}} \simeq c \Vert \vec{k}
\Vert
 \Longrightarrow a_{2x} \simeq \frac{a_0}{z_{2x}} 
 = \frac{a_0^2 H_0}{\sqrt{z_{\rm eq}} c \Vert \vec{k} \Vert} 
\,. \label{a1}
\end{equation}

We have so far discussed only the fate of virtual
particles that happen to emerge from the vacuum. The amount of
particle production also depends upon the rate at which they
emerge. Most types of massless particles possess a symmetry that
makes this rate decrease as the universe expands. So any of these
particles which emerge with $c \Vert \vec{k}_{\rm ph} 
\Vert \ltwid H_I$ can persist forever without violating the
uncertainty principle, but not many emerge. 

This symmetry is called conformal invariance. Particle
physicists find it convenient to describe symmetries, and all the
other properties of a dynamical system, in terms of the system's
 Lagrangian. In the classical mechanics of point particles
the Lagrangian is the difference between the kinetic and potential
energies, but the concept can be generalized to describe any sort
of system, including the quantum field theories responsible for the
vacuum polarization effect described in Sec.~\ref{Vacuum
polarization in flat space}. A field theory is said to be
conformally invariant if its Lagrangian density (the Lagrangian is
the spatial integral of the Lagrangian density) is unchanged when
we multiply each field by a certain power of an arbitrary function
of space and time
$\Omega(x)$. Some interesting fields are the metric $g_{\mu\nu}$,
the vector potential $A_{\mu}$, the Dirac field $\psi_b$ (with $b
= 1,2,3,4$) of spin $\frac12$ particles, and the scalar field
$\phi$. Their conformal transformations are,
\begin{equation}
g_{\mu\nu} \rightarrow \Omega^2 g_{\mu\nu}, \quad A_{\mu}
\rightarrow A_{\mu}, \quad \psi_b \rightarrow \Omega^{-\frac32}
\psi_b, \quad \phi
\rightarrow 
\Omega^{-1} \phi \,. \label{ctrans}
\end{equation}
A typical conformally invariant Lagrangian density is that of electromagnetism,
\begin{equation}
{\cal L}_{EM} = -\frac 14 F_{\alpha\beta} F_{\rho\sigma} g^{\alpha\rho} 
g^{\beta\sigma} \sqrt{-g}, \label{EM Lagrangian}
\end{equation}
where $g = {\rm det}(g_{\mu\nu})$ denotes the determinant of the metric and
$g^{\mu\nu}$ is its matrix inverse.

Conformal invariance is so important because there is a coordinate system 
in which a general homogeneous and isotropic metric,
Eq.~(\ref{HI}), is just a conformal factor times the metric of
flat space. The change of variables is defined by the differential
relation, $d\eta = dt/a(t)$,
\begin{equation}
ds^2 = -dt^2 + a^2(t) d\vec{x} \cdot d\vec{x} = a^2 [- d\eta^2 +
d\vec{x} 
\cdot d\vec{x} ] \,. \label{conf}
\end{equation}
In the $(\eta,\vec{x})$ coordinate system --- which we shall
henceforth employ, except as noted --- the metric and its inverse
are,
\begin{equation}
g_{\mu\nu} = a^2 \eta_{\mu\nu}, \qquad g^{\mu\nu} = a^{-2}
\eta^{\mu\nu}
\,, \label{confgmn}
\end{equation}
where $\eta_{\mu\nu} = \eta^{\mu\nu} = {\rm diag}(-1,1,1,1)$. In
this coordinate system, a conformally invariant Lagrangian density
is the same as in flat space when expressed in terms of the
conformally rescaled fields Eq.~(\ref{ctrans}) with $\Omega =
a^{-1}$. For example, the Lagrangian densities of electromagnetism,
massless Dirac fermions, and a massless, conformally coupled,
complex scalar are,
\begin{eqnarray}
{\cal L}_{EM} & = & -\frac14 F_{\alpha\beta} F_{\rho\sigma} 
\eta^{\alpha\rho} \eta^{\beta\sigma} \,, \label{EM Lagrangian 2}
\\ {\cal L}_{D} & = & i (a^{\frac32} {\overline \psi}_b)
\gamma_{bc}^{\mu} 
\partial_{\mu} (a^{\frac32} \psi_c) \,, \label{Dirac} \\
{\cal L}_{CS} & = & -\partial_{\mu} (a \phi^*) \partial_{\nu} (a \phi) 
\eta^{\mu\nu} \,. \label{confcoup}
\end{eqnarray}
where $\gamma^{\mu}_{bc}$ (with $b,c = 1,2,3,4$) are the gamma
matrices of Dirac theory~\cite{PeskinSchroeder:1995}.

Although the formalism of Lagrangian quantum field theory is a
straightforward generalization of quantum mechanics, the reader who
is unfamiliar with it need not fret over how Lagrangians and
Lagrangian densities encapsulate the dynamics of any particular
system; it is sufficient for our discussion to note that they do.
Therefore, any two systems whose Lagrangians agree have the
same dynamics. We shall exploit this fact at several points in
the subsequent discussion, starting now. Because the massless
particle Lagrangian densities in Eqs.~(\ref{EM Lagrangian
2})--(\ref{confcoup}) agree for the conformally rescaled fields
with those of flat space, it follows that the rate at which virtual
particles emerge from the vacuum is the same in conformal
coordinates as it is in flat space. Let us call this constant rate
$\Gamma$. It gives the number of virtual particles emerging 
per unit conformal time $\eta$. This means that the number per
unit physical time $t$ is,
\begin{equation}
\frac{dn}{dt} = \frac{d\eta}{dt} \frac{dn}{d\eta} =
\frac{\Gamma}{a} \,.
\label{smallrate}
\end{equation}
Hence the emergence rate falls like $1/a(t)$ in physical time, and
we see that the production of massless, conformally invariant
particles is highly suppressed.

Readers who have studied undergraduate quantum mechanics are
familiar with the minimal coupling prescription by which one
generalizes the Schr\"odinger equation to account for the
interaction of a point particle with an electromagnetic field. 
What one does is to replace the derivative operator
$\partial_{\mu}$ everywhere with the covariant derivative
$\partial_{\mu} + i (e/\hbar c) A_{\mu}(x)$, where $e$ is the particle's
charge and $A_{\mu}(x)$ is the 4-vector potential evaluated at the
particle's position in spacetime. A very similar procedure exists
for generalizing the dynamics of any system from flat space to
curved space. This prescription is also called minimal coupling.
For electromagnetism and massless Dirac fermions it happens to 
produce the conformally invariant Lagrangian densities Eqs.~(\ref{EM
Lagrangian 2}) and (\ref{Dirac}). However, when a massless scalar
is minimally coupled, the Lagrangian density which results is not
Eq.~(\ref{confcoup}), but rather,
\begin{equation}
{\cal L}_{MS} = - \partial_{\mu} \phi^* \partial_{\nu} \phi
g^{\mu\nu}
\sqrt{-g} = -a^2 \partial_{\mu} \phi^* \partial_{\nu} \phi
\eta^{\mu\nu}.
\label{scalar Lagrangian}
\end{equation}
This Lagrangian density is not conformally invariant, which means
that the rate at which massless, minimally coupled scalars emerge
from the vacuum per unit physical time need not fall off like the
rates (\ref{smallrate}) of other massless particles. We shall now
show, by direct calculation, that it does not.

Recall that the Lagrangian of any field theory is obtained by
integrating the Lagrangian density over space. Doing this in the
original $(t,\vec{x})$ coordinate system for Eq.~(\ref{scalar
Lagrangian}) and applying Parseval's theorem gives
\begin{eqnarray}
L_{MS} &=& \int d^3x {\cal L}_{MS} 
 \label{SHO}
\\
 &=& \frac{a^3(t)}{c^2} \int \!\! \frac{d^3k}{(2\pi)^3} \Bigl\{
\vert \widetilde{\dot{\phi}}(t,\vec{k}) \vert^2 - \frac{\Vert c \vec{k} 
\Vert^2}{a^2(t)} \vert \widetilde{\phi}(t,\vec{k}) \vert^2 \Bigr\}.
\quad
\nonumber
\end{eqnarray}
Now note that integration is a form of summation. The correspondence can be 
made explicit by changing to dimensionless variables $\vec{k} = 2 \pi \vec{n}
H_I/c$, and then exploiting the Maclaurin relation between sums and
integrals (familiar from undergraduate statistical mechanics),
\begin{equation}
\int \!\! \frac{d^3k}{(2\pi)^3} = \Bigl(\frac{H_I}{c}\Bigr)^3 \!\int
d^3n 
\longrightarrow \Bigl(\frac{H_I}{c}\Bigr)^3 \sum_{\vec{n}} \,.
\end{equation}
We can define point particle positions using the real and imaginary
parts of 
$\widetilde{\phi}(t,\vec{k})$,
\begin{equation}
\widetilde{\phi}(t,\vec{k}) \equiv \sqrt{\frac{\hbar c^3}{2 H_I^2}} \Bigl(
x(t,\vec{n}) + i y(t,\vec{n}) \Bigr) \,.
\end{equation}
Hence the scalar Lagrangian,
\begin{equation}
L_{MS} \longrightarrow \frac{\hbar H_I a^3}{2 c^2} \sum_{\vec{n}} \Bigl\{
\dot{x}^2 + \dot{y}^2 - \Bigl(\frac{c \Vert \vec{k} \Vert}{a} \Bigr)^2 (x^2
+ y^2) \Bigr\},
\end{equation}
reveals that each wave vector of the scalar corresponds to two
independent harmonic oscillators with the following time dependent
mass and frequency,
\begin{equation}
m(t) = c^{-2} \hbar H_I a^3(t) \quad {\rm and} \quad \omega(t) = c \Vert 
\vec{k} \Vert/a(t) \,.
\end{equation}

Harmonic oscillators with a time dependent mass and frequency have
been much studied in quantum mechanics~\cite{Gasiorowicz:1974}.
The minimum energy at time
$t$ is well known to be $\frac12 \hbar \omega(t)$, however, the
state with this energy does not generally evolve onto itself. For
the inflationary case of $a(t) \propto e^{H_I t}$, the system's time
dependence can be solved exactly. The state whose energy is minimum
in the distant past has instantaneous average (zero-point) energy,
\begin{equation}
E_{0\!-\!point}(t,\vec{k}) = \frac{\hbar c \Vert \vec{k} \Vert}{a(t)} + 
\frac{\hbar H_I^2 a(t)}{2 c \Vert \vec{k} \Vert} \,. \label{0pt}
\end{equation}
The second term in Eq.~(\ref{0pt}) is attributable to particle
production. The energy of a single particle with this wave vector
is $\hbar c \Vert 
\vec{k} \Vert/a(t)$, so the average number of particles with wave vector
$\vec{k}$ is,
\begin{equation}
N(t,\vec{k}) = \frac12 \Bigl(\frac{H_I a(t)}{c \Vert \vec{k} \Vert}\Bigr)^2
\,.
\end{equation} As we expect, $N(t,\vec{k})$ is small for very early
times and becomes comparable to one at horizon crossing. If we sum
the contributions from all 
wave vectors that have experienced horizon crossing and divide by
the spatial volume, we find the number density,
\begin{equation}
\frac{N}{V} = \frac{H_I^3}{4 \pi^2 c^3} \,.
\end{equation}
This corresponds to $1/8\pi^2$ particles per Hubble volume for each
degree of freedom. 

We close by commenting that there can be no question about the reality of 
inflationary particle production because its impact has been detected. There
is strong evidence that it is what caused the anisotropies imaged by
WMAP~\cite{Bennett:wmap2003}. Indeed, all the cosmological
structures of the current universe are the result of gravitational
collapse into these (originally) quantum fluctuations over the
course of 13.7 billion years!

\section{Vacuum polarization in inflation}
\label{Vacuum polarization in inflation}

The inflationary Hubble parameter Eq.~(\ref{infHub}) corresponds
to an enormous energy, 
\begin{equation}
\hbar H_I \simeq 10^{3}~{\rm J} \simeq 10^{13}\,{\rm GeV} \,.
\label{infE}
\end{equation}
On this scale all the charged particles in the Standard Model of
particle physics are effectively massless. Even a particle we
normally consider very massive, such as the $t$ quark, has less
than $10^{-10}$ times as much rest mass energy. However, all but
one of the Standard Model charged particles are described by the
Dirac field $\psi_a$, whose Lagrangian density, Eq.~(\ref{Dirac}),
becomes conformally invariant when we ignore masses. As explained
in Sec.~\ref{Virtual particles with expansion}, conformally
invariant particles are not produced much during inflation. This
means that they do not contribute much more to the polarization of
the vacuum during inflation than they do in flat space.

The lone exception within the Standard Model of elementary particle theory
is the charged sector of the Higgs scalar. At low energy it manifests as the 
longitudinal component of the $W^{\pm}$. Its mass of about 80\,GeV
is also insignificant on the scale of inflation. No one really
knows how it couples to the metric, but the usual assumption, based
on how the field renormalizes, is minimal coupling. We can
therefore model it with the Lagrangian density of a massless,
charged and minimally coupled scalar,
\begin{eqnarray}
{\cal L}_{SQED} \! & = & \! -(\partial_{\mu} - ie' A_{\mu}) \phi^* 
(\partial_{\nu} + i e' A_{\nu} ) \phi g^{\mu\nu} \sqrt{-g}
\quad\nonumber\\
\! & = & \! -a^2 (\partial_{\mu} - ie' A_{\mu}) \phi^* (\partial_{\nu} + 
i e' A_{\nu} ) \phi \eta^{\mu\nu}. \label{SQEDL}
\end{eqnarray}
(Here $e' \equiv e/\hbar c$, and $e \simeq -0.30 \sqrt{\hbar c}$ is
the charge of the electron. The subscript SQED stands for 
scalar quantum electrodynamics.) There may be more so-far
undiscovered charged scalars of this type lurking between the
$\sim 10^2$\,GeV energies which can be explored at
accelerators and the enormous energy in Eq.~(\ref{infE}) of the 
inflationary Hubble parameter.

With Ola T\"ornkvist we have computed the vacuum polarization from
${\cal L}_{
SQED}$~\cite{ProkopecTornkvistWoodard:2002,ProkopecTornkvistWoodard:2002b}.
In the $x^{\mu} = (\eta,\vec{x})$ coordinates Eq.~(\ref{conf}) our
result for the polarization bi-tensor takes the same form
Eq.~(\ref{genPi}) as it does for the linear, isotropic medium
discussed in Sec.~\ref{Vacuum polarization in flat space}. With
the scale factor normalized to unity at the start of inflation the
two scalar functions are,
\begin{eqnarray}
\chi_e(x,x') &=& \chi_e^{\rm flat}\!(x,x') - \frac{\alpha}{6\pi}
\ln(a) \delta^4(x-x') 
\nonumber\\
\;+\, \frac{\alpha a a' H_I^2}{8 \pi^2 c^2} 
\!\!\!\!\!&&\!\!\!\!\!\!
 \partial^2 \Bigl\{ 
\theta(\Delta \eta) \theta(\Delta \tau^2) [1 + \ln(H_I^2 {\Delta 
\tau}^2)] \Bigr\}
\quad
 \label{SQED1} 
\\
\delta n^2(x,x') &=& \frac{\alpha a^2 a^{\prime 2} H_I^4}{4 \pi^2
c^4} \theta(\Delta \eta) \theta({\Delta \tau}^2) [2 + \ln(H_I^2
{\Delta 
\tau}^2)] 
\nonumber\\
\label{SQED2}
\end{eqnarray}
Here ${\Delta \eta} \equiv \eta - \eta'$ and $\Delta \tau^2 \equiv
(\eta -
\eta')^2 - c^{-2} \Vert \vec{x} - \vec{x}' \Vert^2$. 
$\chi_e^{\rm flat}\!(x,x')$ is the flat space result~(\ref{SQED}), 
with $t$ and $t'$ 
replaced by $\eta$ and $\eta'$. This term is renormalized precisely as in 
flat space, and contains no scale factors. The inflationary
corrections are completely finite and depend on the scale
factors $a \equiv a(\eta)$ and $a' \equiv a(\eta')$. These
inflationary corrections come from the long wave length virtual
particles that are ripped out of the vacuum by the inflationary
Hubble flow. This should obviously increase polarization because 
it fills spacetime with a plasma of charged particles.

A significant feature of our result is nonzero $\delta n^2(x,x')$. Recall 
that it must always vanish in flat space quantum field theory by virtue of 
the Lorentz symmetry between space and time. The time-dependent metric of 
inflation, Eq.~(\ref{confgmn}), does not possess this symmetry, so
$\delta n^2(x,x') \neq 0$. In terms of electrodynamics, this means that 
${\cal L}_{SQED}$ induces a relative permittivity which is not the 
inverse of the permeability, so the index of refraction is not unity even in 
``empty'' space.

Because the inflationary metric is time dependent, we cannot
calculate the mass of the photon by checking for a pole in the
Fourier transform of the susceptibility as we did in flat space,
Eq.~(\ref{howto}). A better way to proceed is by comparison with
the Proca Lagrangian density which governs the dynamics of a
fundamental massive photon,
\begin{eqnarray}
{\cal L}_{\rm P} \! & \equiv & \! -\frac14 F_{\alpha \beta} F_{\rho 
\sigma} g^{\alpha \rho} g^{\beta\sigma} \! \sqrt{-g} - \frac{m_{\gamma}^2 
c^2}{2 \hbar^2} A_{\mu} A_{\nu} g^{\mu \nu} \! \sqrt{-g}
\nonumber\\
\! & = & \! -\frac14 F_{\alpha \beta} F_{\rho \sigma} \eta^{\alpha
\rho} 
\eta^{\beta\sigma} - \frac{m_{\gamma}^2 c^2}{2 \hbar^2} A_{\mu}
A_{\nu} 
\eta^{\mu \nu} a^2. 
\label{ProcaL}
\end{eqnarray}
The field equations associated with this Lagrangian density are,
\begin{equation}
\eta^{\mu\nu} \eta^{\rho\sigma} \partial_{\rho} F_{\sigma\nu} - c^2
\hbar^{\!-2} m_{\gamma}^2 \eta^{\mu\nu} A_{\nu} a^2 = 0 \,.
\label{Procaeqn}
\end{equation}
The mass term is distinguished by its factor of $a^2$.

Now recall Maxwell's equations with vacuum polarization,
Eq.~(\ref{genMax}), which we rewrite without the current,
\begin{equation}
\eta^{\mu\nu} \eta^{\rho\sigma} \partial_{\rho} F_{\sigma\nu}(x) 
+ \!
\int \! d^4x' [\mbox{}^{\mu} \Pi^{\nu}]\!(x,x') A_{\nu}(x') = 0.
\label{Max+pol}
\end{equation}
We also recall the polarization bi-tensor Eq.~(\ref{genPi}),
\begin{eqnarray}
[^{\mu} \Pi^{\nu}]\!(x,x') &\equiv& - [\eta^{\mu\nu} \eta^{\rho\sigma}
- \eta^{\mu\rho} \eta^{\sigma\nu}] \partial^{\prime}_{\rho} \partial_{\sigma}
\chi_e(x,x') 
\nonumber\\
&+& \eta^{\mu i} \eta^{\nu j} [\delta_{ij}
\partial^{\prime}_k 
\partial_k - \partial^{\prime}_i \partial_j] \delta n^2(x,x') \,. 
\quad
\end{eqnarray}
We see from Eq.~(\ref{SQED2}) that $\delta n^2(x,x')$
contributes a factor of 
$a^2$. The $\chi_e(x,x')$ term, Eq.~(\ref{SQED1}), has at most a
single factor of $a$, but note from $\partial_{\sigma} a =
\delta^0_{\sigma} H_I a^2$ that this term can also give an
$a^2$ in Eq.~(\ref{Max+pol}). A comparison with the Proca equations
(\ref{Procaeqn}) suggests $m_{\gamma} \sim \sqrt{\alpha} H_I 
\hbar/c^2$.

We can obtain a quantitative result by solving Eq.~(\ref{Max+pol})
perturbatively in $\alpha$. First expand the vector potential in
a series of terms 
$A_{\mu}^{(n)}\!(x)$ which go like $\alpha^n$,
\begin{equation}
A_{\mu}(x) = A_{\mu}^{(0)}\!(x) + A_{\mu}^{(1)}\!(x) + \dots 
\end{equation}
Now recall that the polarization bi-tensor is first order in
$\alpha$, and it groups the terms in Eq.~(\ref{Max+pol}) in powers of
$\alpha$. We see that 
$A_{\mu}^{(0)}\!(x)$ obeys the classical equation,
\begin{equation}
\eta^{\rho\sigma} \partial_{\rho} F^{(0)}_{\sigma\nu}\!(x) = 0,
\end{equation}
the general solution of which consists, in the Coulomb gauge, of a
superposition of transverse plane waves,
\begin{equation}
A_{\mu}^{(0)}\!(x) = \epsilon_{\mu}(\vec{k}) e^{-i c \Vert \vec{k} \Vert \eta
+ i \vec{k} \cdot \vec{x}} \,,\quad
 {\rm where}\ \epsilon_0 = 0 = k_i \epsilon_i.
\quad
\label{plane}
\end{equation}
The order $\alpha$ correction obeys,
\begin{equation}
\eta^{\mu\nu} \eta^{\rho\sigma} \partial_{\rho}
F^{(1)}_{\sigma\nu}\!(x) \! + 
\! \int \! d^4x' [\mbox{}^{\mu} \Pi^{\nu}] (x,x')
A^{(0)}_{\nu}\!(x') = 0.
\end{equation}
Now substitute Eq.~(\ref{plane}) and evaluate the integral
assuming the photon experienced first horizon crossing (see
Fig.~\ref{figure-III}) long before, and after a long period of
inflation, 
\begin{equation}
a \gg \frac{c \Vert \vec{k} \Vert}{H_I} \gg 1 \,.
\end{equation}
After some tedious expansions, the result is,
\begin{eqnarray}
&&\int \! d^4x' [^{\mu} \Pi^{\nu}](x,x') A^{(0)}_{\nu} (x') 
\\
\!&=&\!\! 
 -\alpha c^{-2} H_I^2 \eta^{\mu\nu} A_{\nu}^{(0)} (x) 
\Bigl[ \frac{2}{\pi} \ln \Bigl(\frac{c \Vert \vec{k}
\Vert}{H_I}\!\Bigr) 
 + O(1)\Bigr] a^2 + O(a). 
\nonumber
\end{eqnarray}
The analogous first order Proca equation,
\begin{equation}
\eta^{\mu\nu} \eta^{\rho\sigma} \partial_{\rho} F^{(1)}_{\sigma\nu} - 
c^2 \hbar^{\!-\!2} m_{\gamma}^2 \eta^{\mu\nu} A^{(0)}_{\nu} a^2 = 0
\,,
\end{equation}
implies that the photon mass must be,
\begin{equation}
m_{\gamma} = \sqrt{\alpha} c^{-\!2} \hbar H_I \Bigl[\frac{2}{\pi} 
\ln\Bigl( \frac{c \Vert \vec{k} \Vert}{H_I}\!\Bigr)+ O(1) 
\Bigr]^{\frac12}. \label{photon mass}
\end{equation}
We have thus discovered that the photon of scalar electrodynamics
acquires mass during inflation.

\section{Hartree approximation}
\label{Hartree approximation}

A simple way of obtaining almost the same result was previously
suggested by one of us (Prokopec) in collaboration with
Anne-Christine Davis, Konstantinos Dimopoulos, and Ola
T\"ornkvist~\cite{DavisDimopoulosProkopecTornkvist:2000,
TornkvistDavisDimopoulosProkopec:2000,DimopoulosProkopecTornkvistDavis:2001}.
The technique is to pretend that photons move in the quantum
mechanical average of the scalar field. This is known as the
Hartree or mean field approximation. First year graduate students
should be familiar with its use in quantum mechanics to treat
interacting, multi-electron atoms, and in statistical mechanics to
treat interacting, multi-particle systems such as gases subject to
the van der Waals force.

To implement the Hartree approximation we first take the average of 
${\cal L}_{SQED}$ in Eq.~(\ref{SQEDL}) over quantum mechanical
fluctuations of the scalar field. Of course this average eliminates
the scalar fields, but it leaves behind some function of the vector
potential,
\begin{eqnarray}
\Bigl\langle {\cal L}_{SQED} \Bigr\rangle &=& - \Bigl\langle 
\partial_{\mu} \phi^* \partial_{\nu} \phi \Bigr\rangle \eta^{\mu\nu} a^2 
\nonumber\\
&+&
\frac{i e}{\hbar c} \Bigl\langle \phi^* \partial_{\mu} \phi - \partial_{\mu} 
\phi^* \phi \Bigr\rangle 
A_{\nu} \eta^{\mu\nu} a^2 
\nonumber\\
&-& \frac{e^2}{\hbar^2 c^2} 
\Bigl\langle \phi^* \phi^* \Bigr\rangle A_{\mu} A_{\nu} \eta^{\mu\nu} a^2. 
 \label{VEV}
\end{eqnarray}
Now add this function to the Maxwell Lagrangian density
Eq.~(\ref{EM Lagrangian}).

By using the sophisticated techniques of quantum field theory, we
can show that the quantum average of the scalar's norm-squared
consists of a divergent constant plus a finite term that grows
like the logarithm of the scale factor~\cite{scalarvev}:
\begin{equation}
\Bigl\langle \phi^*(x) \phi(x) \Bigr\rangle = {\rm U.V.} + \frac{H_I^2 \hbar}{
4\pi^2 c} \ln(a) \,. \label{mean}
\end{equation}
The other averages in Eq.~(\ref{VEV}) are either zero or else
they do not multiply functions of the vector potential. The
Hartree approximation Lagrangian density is therefore, 
\begin{eqnarray}
{\cal L}_{\rm Hartree} &=& -\frac 14 F_{\alpha\beta}
F_{\rho\sigma} 
\eta^{\alpha\rho} \eta^{\beta\sigma} + {\rm U.V.} 
\nonumber\\
&-& \frac{e^2}{\hbar^2 c^2} \Bigl[{\rm U.V.} 
+ \frac{H_I^2 
\hbar}{4\pi^2 c} \ln(a) \Bigr] A_{\mu} A_{\nu} \eta^{\mu\nu} a^2.
\qquad
\label{Hartree}
\end{eqnarray}

Expression~(\ref{Hartree}) contains ultraviolet divergences
because virtual particles of arbitrarily large wave vector
contribute to the average. This is the same origin as the
divergences we found in Sec.~\ref{Vacuum polarization in flat
space}, and it is the ultimate origin of all ultraviolet
divergences in quantum field theory. The only thing that need
concern us here is the dependence of the divergent terms on the
electromagnetic vector potential 
$A_{\mu}(x)$. The divergence without any vector potentials is harmless, but 
the other one could only be renormalized using a fundamental photon mass, 
which we do not have. This is one reason why the vacuum polarization --- 
which can be consistently renormalized --- is the correct way to
study the kinematical properties of photons. But let us simply
ignore the divergences in Eq.~(\ref{Hartree}). A comparison of
the finite parts with the Proca Lagrangian density
Eq.~(\ref{ProcaL}) suggests the correspondence,
\begin{equation}
m_{\gamma} \Longleftrightarrow \sqrt{\alpha} c^{\!-\!2} \hbar H_I 
\Bigl[\frac2{\pi} \ln(a) \Bigr]^{\frac12} \,. \label{timedep}
\end{equation}
Complete agreement with Eq.~(\ref{photon mass}) requires only the
additional assumption that the growth of Eq.~(\ref{timedep})
ceases for the mode of wave vector $\vec{k}$ (which we assume
obeys
$\Vert \vec{k} \Vert \gg H_I/c$) when it experiences horizon
crossing, $a = c \Vert \vec{k} \Vert/H_I$ (see
Fig.~\ref{figure-III}). This point of view is consistent with
the causal picture according to which a photon's mass only receives
contributions from virtual scalars whose wave lengths are
greater than the photon's wave length.

We conclude this section by commenting on the size of the photon
mass induced by our mechanism. During inflation we have $m_\gamma
\sim 10^{13}\,{\rm GeV}/c^2$, which is enormous compared to the
center-of-mass energies of $\sim 100\,{\rm GeV}$ attainable in the
largest accelerators. A photon mass is not detected today
because our result derives from the huge density of free charged
particles ripped out of the vacuum by inflation. This plasma has
been thoroughly dissipated at any wave length we can access in
today's laboratories.

According to the supernovae
results~\cite{Perlmutter-etal:1998,Riess-etal:1998}, the current
universe may be entering another phase of inflation. This late inflationary
phase also leads to a nonzero photon mass, but with the
replacement of
$H_I$ by the vastly smaller Hubble parameter of today, $H_0$.
This substitution in Eq.~(\ref{photon mass}) gives a
minuscule photon mass, $m_\gamma
\sim 10^{-42}\,{
\rm GeV}/c^2$, which is far below the best current laboratory bounds
of 
$m_\gamma \lsim 10^{-49}\,{\rm kg} \approx 10^{-23}\,{\rm
GeV/c^2}$~\cite{WilliamsFallerHill:1971,GoldhaberNieto:1971}.

\section{Cosmological magnetic fields}
\label{Cosmological magnetic fields}

The phenomenon of nonzero photon mass during inflation offers a
fascinating 4-dimensional analogue to the Schwinger model in two
dimensions~\cite{Schwinger:1962}. However, it was
proposed~\cite{DavisDimopoulosProkopecTornkvist:2000,
TornkvistDavisDimopoulosProkopec:2000,DimopoulosProkopecTornkvistDavis:2001}
not for aesthetic appeal, but rather to explain the curious fact
that galaxies seem to possess magnetic fields,
which are correlated on scales of a few kilo-parsecs,
and whose strength is typically a few micro-Gauss~\cite{Kronberg:1994}.
(The conversion factors to MKSA units are 
$1\,{\rm J/m}^3 = 10\,{\rm Gauss}^2$ and $10\,{\rm kpc} \simeq 3.1
\times 10^{20}\,{\rm m}$.) There is also evidence that galactic
clusters possess micro-Gauss magnetic fields correlated on scales
of 10--100\,kpc~\cite{GrassoRubinstein:2001}. Although some of the
material in this section is more technical than before, we present
it to illustrate how the phenomenon of vacuum polarization during
inflation may have left an observable consequence.

The difficulty with cosmic magnetic fields is not their field
strengths but rather their enormous coherence lengths. A galaxy's
differential rotation can combine with the turbulent motion of
ionized gas to power a phenomenon known as the $\alpha$-$\omega$
dynamo~\cite{dynamo}. In this mechanism the lines of a
coherent seed field are stretched by rotation, twisted by 
turbulence, and then recombined to result in an exponential
amplification, 
$B(t) \propto e^{t/\tau}$. Kinetic energy from the turbulent motion
is converted into magnetic field energy in this way until
equipartition is reached. Although many astrophysicists question
the $\alpha$-$\omega$ dynamo, it is significant that the measured
field strengths are at roughly the equipartition 
limit~\cite{GrassoRubinstein:2001}.

There is no general agreement on a reasonable value for the dynamo time
constant $\tau$; estimates vary from 0.2 to 0.8 billion
years~\cite{DavisLilleyTornkvist:1999}.
By observing a surprisingly large polarization in the cosmic microwave 
background photons, the WMAP satellite has seen
reionization from the first star formation at about 0.2 billion
years into the 13.7 billion years of the universe's
existence~\cite{Spergel:wmap2003}. One might expect large spiral
galaxies to form at about 0.4 billion
years~\cite{DavisLilleyTornkvist:1999}, which implies dynamo operation
for $13.7 - 0.4 = 13.3$ billion years, or between 17 to 66 time
constants. Exponentiation results in amplification factors ranging
from $e^{17} \simeq 2.4 \times 10^7$ to as large as $e^{66} \simeq
4.6 \times 10^{28}$. Therefore the cosmological magnetic fields of
today might derive from correlated seeds as weak as 
$10^{-34}\,{\rm Gauss}$ at the time of galaxy formation. The real
question is what produced the correlated seed fields in the hot,
dense and very smooth early universe?

In the following, we argue that the nonzero photon mass of inflation
might help to answer this question. 
As explained in Sec.~\ref{Virtual particles with
expansion}, a nonzero mass suppresses the creation of particles. On
the other hand, it vastly enhances the zero-point energy that
quantum mechanics predicts must reside in each photon wave vector 
$\vec{k}$, even if there are no particles with that wave vector
anywhere in the universe. With no mass this zero-point energy falls
as the universe expands,
\begin{equation}
E_{\gamma}(t,\vec{k})\Bigl\vert_{m_{\gamma} = 0} = \frac{\hbar c
\Vert 
\vec{k} \Vert}{2 a(t)} \,.
\end{equation}
A nonzero photon mass causes the zero-point energy of wave vectors
that have experienced first horizon crossing to approach a
constant instead,
\begin{equation}
E_{\gamma}(t,\vec{k}) = \frac12 \sqrt{m_{\gamma}^2 c^4 + \hbar^2
c^2 \Vert
\vec{k} \Vert^2/a^2} \longrightarrow \frac12 m_{\gamma} c^2 \,.
\end{equation}
After the end of inflation this wave vector eventually experiences second
horizon crossing, $c \Vert \vec{k} \Vert \simeq a(t) H(t)$. If the mass goes
to zero quickly thereafter, about half of the enormous zero-point
energy must be shed in the form of long wave length photons at
numbers vastly higher than thermal. The idea is that the
mysterious seed fields derive from these long wave length photons
becoming frozen in the plasma of the early universe.

Consider a wave vector $\vec{k}$ that is about to experience second
horizon crossing. Each polarization of this system behaves as an
independent harmonic oscillator whose frequency is suddenly
changed from a large value $\Omega = m_{\gamma} c^2/\hbar$ to a
much smaller one $\omega = c \Vert \vec{k} 
\Vert/a(t)$. Let $q$ and $p$ stand for the position and momentum operators of 
this oscillator. The Hamiltonians before and after are,
\begin{equation}
H_B = \frac{p^2}{2 m} + \frac12 m \Omega^2 q^2, \qquad
H_A = \frac{p^2}{2 m} + \frac12 m \omega^2 q^2 \,.
\end{equation}
Before transition the system is in its ground state,
\begin{equation}
H_B \vert 0 \rangle = \frac12 \hbar \Omega \vert 0 \rangle \,.
\end{equation}
This result can be obtained from the standard relations, 
\begin{eqnarray}
q = \sqrt{\frac{\hbar}{2m\Omega}}\Bigl(\widehat{a} -
\widehat{a}^\dagger\Bigr)
\,,\quad
 p = \sqrt{\frac{\hbar m \Omega}{2}}\Bigl(-i \widehat{a} + i 
\widehat{a}^\dagger\Bigr), 
\quad
\label{q-p}
\end{eqnarray}
where $\widehat{a} = \widehat{a}_0 \exp(-i\Omega t)$ and
$\widehat{a}^\dagger = \widehat{a}_0^\dagger \exp(i\Omega t)$ are
the lowering and raising operators, respectively, with
$\widehat{a}_0| 0\rangle = \widehat{a}| 0\rangle = 0$, not to be
confused with the scale factor $a(t)$. The kinetic and potential
terms each contribute half,
\begin{equation}
\frac1{2m} \langle 0 \vert p^2 \vert 0 \rangle = \frac14 \hbar \Omega =
\frac12 m \Omega^2 \langle 0 \vert q^2 \vert 0 \rangle \,.
\end{equation}
After the transition the system is no longer an eigenstate, but we
can find its average energy from the fact that the expectation
values of $p^2$ and 
$q^2$ are continuous,
\begin{equation}
\langle 0 \vert H_A \vert 0 \rangle = \frac14 \hbar \Omega \Bigl[1 + \frac{
\omega^2}{\Omega^2}\Bigr] \approx \frac14 \hbar \Omega \,.
\end{equation}
If we compare this energy with the post-transition eigenstates
$(N+\frac12) \hbar 
\omega$, we see that the average occupation number after transition
is,
\begin{equation}
N\!(\vec{k}) \simeq \frac{\Omega}{4 \omega} = \frac{m_{\gamma} c a}{4 \hbar 
\Vert \vec{k} \Vert} \,. \label{N1}
\end{equation}
The substitution in Eq.~(\ref{N1}) of the scale factor at the second
horizon crossing, 
$a=a_{2x}$ as given in Eq.~(\ref{a1}), results in the average
occupation number for each polarization of wave vector $\vec{k}$,
\begin{equation}
N\!(\vec{k}) = \frac{m_{\gamma} H_0 a_0^2}{4 \sqrt{z_{\rm eq}}
\hbar \Vert \vec{k} 
\Vert^2} \,. \label{N2}
\end{equation}
(Recall that $k_{\rm ph} = \Vert \vec{k}\Vert/a_0$ is the physical wave
number measured today.) To within factors of order unity, the
temperature at time $t$ is $T = T_0 a_0/a(t)$, where $T_0 \simeq
2.73\,{\rm K}$ is the current temperature of the cosmic microwave
background. The number of thermal photons of wave number $\vec{k}$
and a fixed polarization obeys the Planck distribution,
\begin{equation}
N_{\rm th}\!(\vec{k}) = \Bigl[\exp\
 \Bigl({\frac{\hbar c \Vert \vec{k} \Vert}{k_B T_0 a_0}}\Bigr)
- 1 \Bigr]^{-1} 
\,\stackrel{IR}{\longrightarrow}\; \frac{k_B T_0 a_0}{\hbar c
\Vert \vec{k} 
\Vert} \,, \label{Nth}
\end{equation}
where $k_B \simeq 1.38 \times 10^{-23}\,{\rm J/K}$ is the Boltzmann
constant and the final relation on the right applies in the long
wave length (IR) limit. The ratio of $m_{\gamma}$ photons to
thermal ones can be expressed in terms of the present-day physical
wave length $\lambda_0$,
\begin{eqnarray}
\frac{N(\vec{k})}{N_{\rm th}(\vec{k})} &\simeq& 
\frac{m_{\gamma} c H_0 
\lambda_0}{8 \pi \sqrt{z_{\rm eq}} k_B T_0} 
\nonumber\\
&=& \frac{\hbar
H_I H_0 
\lambda_0}{8 \pi \sqrt{z_{\rm eq}} c k_B T_0} \Bigl[ \frac{2
\alpha}{\pi} 
\ln\Bigl(\frac{2 \pi c a_0}{H_I \lambda_0} \Bigr) \Bigr]^{\frac12}.
\end{eqnarray}
Recall that we normalize the scale factor to one at the start of
inflation. For models with a long period of inflation the final
factor in square brackets is dominated by $\ln(a_0)$, which might
be quite large. We parameterize our ignorance of the number of
inflationary e-foldings by defining,
\begin{equation}
\varepsilon \equiv \Bigl[\frac{2 \alpha}{\pi} \ln\Bigl(\frac{2 \pi c a_0}{H_I
(10~{\rm kpc})}\Bigr)\Bigr]^{\frac12}.
\end{equation}
If we work out the other numbers, we obtain,
\begin{equation}
\frac{N(\vec{k})}{N_{\rm th}(\vec{k})} \simeq \varepsilon \times
10^{\! -\!4} 
\Bigl( \frac{\lambda_0}{\rm m}\!\Bigr) \simeq \varepsilon \times
10^{17} 
\Bigl( \frac{\lambda_0}{10\,{\rm kpc}} \Bigr).
\end{equation}
We see that $m_{\gamma}$ photons are negligible compared to thermal
ones on the $\lambda_0 \sim 0.002$\,m scale of the cosmic microwave
background, but they are enormously dominant on the $\lambda_0 \sim
10$\,kpc scale relevant to galaxies.

At first the energy in these photons is almost completely electric, but 
Maxwell's equations carry it to the magnetic sector. The physical magnetic 
field in a homogeneous and isotropic geometry is,
\begin{equation}
B^i(t,\vec{x}) = -\frac12 \epsilon^{ijk} F_{jk}(t,\vec{x})/a^2(t)
\,.
\end{equation}
If we assume that half of the energy of the massive photons winds up
in these magnetic fields, we conclude that their spatial Fourier
transforms obey, 
\begin{eqnarray}
\Bigl\langle \widetilde{B}^i(t,\vec{k}) \widetilde{B}^i(t,\vec{q})
\Bigr\rangle & \! = \! & (2\pi)^3 \delta^3(\vec{k} + \vec{q})
N(\vec{k}) \frac{\hbar c \Vert 
\vec{k} \Vert}{a^4(t)}, \\
& \! = \! & (2\pi)^3 \delta^3(\vec{k} 
+ \vec{q}) \frac{\varepsilon\hbar H_I H_0 
a_0^2}{4 \sqrt{z_{\rm eq}} c \Vert \vec{k} \Vert a^4}. 
\qquad
\end{eqnarray}

The quantity of interest is the magnetic field averaged over a region of
coordinate size $\ell = \ell_0/a(t)$,
\begin{eqnarray}
B^i(t,\vec{x};\ell_0) & \equiv & (2\pi \ell^2)^{-\frac32} \!\int
\!\! d^3y e^{-\frac{\Vert \vec{x} - \vec{y} \Vert^2}{2 \ell^2}}
B^i(t,\vec{y}), \qquad \\ & = & \!\int \! \frac{d^3k}{(2\pi)^3}
e^{i
\vec{k} \cdot \vec{x}} e^{-\frac12 
\ell^2 \Vert \vec{k} \Vert^2} \widetilde{B}^i(t,\vec{k}) \,.
\end{eqnarray}
$B^i(t,\vec{x};\ell_0)$ is an operator, but its average is a
number,
\begin{eqnarray}
B^2(t,\ell_0) &\equiv& \Bigl\langle B^i(t,\vec{x};\ell_0)
B^i(t,\vec{x};\ell_0) \Bigr\rangle
\nonumber\\
&=& \frac{\varepsilon \hbar H_I H_0 (1 +
z)^2}{16 
\pi^2 \sqrt{z_{\rm eq}} c \, \ell_0^2}.
\end{eqnarray}
If we plug in the known numbers, we find
\begin{equation}
B(t,\ell_0) \simeq \sqrt{\varepsilon} \times 10^{-33}\,{\rm Gauss} 
\Bigl(\frac{1 + z}{\ell_0/10\,{\rm kpc}}\Bigr). \label{final}
\end{equation}
This value is already within the lower range of conceivable seed
fields. Turbulent evolution might contribute a factor of ten by
transferring power from small 
scales~\cite{DavisDimopoulosProkopecTornkvist:2000}. An additional
factor of 
$(\rho_{\rm gal}/\rho_0)^{\frac23} \simeq 3 \times 10^3$ accrues from field
compression when the proto-galaxy collapses. If we assume
$\varepsilon
\sim 1$ and galaxy formation at $z \sim 10$, we might expect field
strengths of about 
$10^{-28}$\,Gauss at $\ell_0 \sim 10$\,kpc.

It should be emphasized that this is just one of many potential explanations 
for cosmological magnetic fields~\cite{GrassoRubinstein:2001}. This
section's analysis is also highly simplified. We need to better
understand the process through which a given wave vector's mass
dissipates at second horizon crossing. A proper calculation would
also require careful study of the dynamics of the electric and
magnetic fields during the epochs of reheating, radiation
domination, and matter domination. 

\begin{acknowledgements}
It is a pleasure to acknowledge Ola T\"ornkvist's collaboration
in much of the work reported here. We have also profited from
conversations with Anne-Christine Davis, Konstantinos
Dimopoulos, Sasha Dolgov,  Hendrik J. Monkhorst, Glenn Starkman
and Tanmay Vachaspati. This work was partially supported by DOE contract 
DE-FG02-97ER41029 and by the Institute for Fundamental Theory of
the University of Florida.
\end{acknowledgements}


\begin{thebibliography}{99}

\bibitem{Hubble:1929} E. Hubble, ``A relation between distance
and radial velocity among extra-galactic nebulae,''
Proc. Nat. Acad. Sci. USA {\bf 15}, 168--173 (1929). 

\bibitem{Wirtz:1922+1924} C.~W.~Wirtz, ``Notiz zur Radialbewegung
der Spiralnebel,'' Astronomische Nachrichten {\bf 216}, 451
(1922); C.~W.~Wirtz, ``De Sitters Kosmologie und die
Radialbewegungen der Spiralnebel,'' Astronomische Nachrichten {\bf
222}, 21 (1924). 

\bibitem{Perlmutter-etal:1998} S. Perlmutter et al. [Supernova
Cosmology Project Collaboration], ``Measurements of Omega and
Lambda from 42 high-redshift supernovae,'' Astrophys. J. {\bf 517},
565--586 (1999) or arXiv:astro-ph/9812133.

\bibitem{Riess-etal:1998} A. G. Riess {\it et al.} [Supernova
Search Team Collaboration], ``Observational evidence from
supernovae for an accelerating universe and a cosmological
constant,'' Astron. J. {\bf 116}, 1009--1038 (1998) or 
arXiv:astro-ph/9805201.

\bibitem{Bennett:wmap2003} C. L. Bennett {\it et al.}, ``First year
Wilkinson microwave anisotropy probe (WMAP) observations:
Preliminary maps and basic results,'' arXiv:astro-ph/0302207.

\bibitem{Spergel:wmap2003} D. N. Spergel {\it et al.}, ``First year
Wilkinson mi\-cro\-wave an\-iso\-tropy probe (WMAP) observations:
Determination of cosmological parameters,'' arXiv:astro-ph/0302209.

\bibitem{RatraPeebles:1987}
B.~Ratra and P.~J.~Peebles,
``Cosmological consequences of a rolling homogeneous scalar
field,'' Phys. Rev. D {\bf 37}, 3406--3427 (1988).

\bibitem{CaldwellDaveSteinhardt:1997}
R.~R.~Caldwell, R.~Dave and P.~J.~Steinhardt,
``Cosmological imprint of an energy component with general
equation-of-state,'' Phys. Rev. Lett. {\bf 80}, 1582--1585
(1998) or arXiv:astro-ph/9708069.

\bibitem{Guth:1980} A. H. Guth, ``The inflationary universe: A
possible solution to the horizon and flatness problems,'' Phys.
Rev. D {\bf 23}, 347--356 (1981).

\bibitem{Starobinsky:1980} A. A. Starobinsky, ``A new type of
isotropic cosmological models without singularity,'' Phys. Lett. B
{\bf 91}, 99--102 (1980).

\bibitem{Sato:1981} K. Sato, ``Cosmological baryon number domain
structure and the first order phase transition of a vacuum,''
Phys. Lett. B {\bf 99}, 66--70 (1981).

\bibitem{Kazanas:1980} D. Kazanas, ``Dynamics of the universe and
spontaneous symmetry breaking,'' Astrophys. J. {\bf 241}, L59--L63
(1980).

\bibitem{Jackson:III} J. D. Jackson, {\sl Classical
Electrodynamics} (John Wiley \& Sons, New York, 1999), 3rd
ed., Chap. 7.

\bibitem{PeskinSchroeder:1995} M. E. Peskin and D. V.
Schroeder, {\it An Introduction to Quantum Field Theory}
(Addison-Wesley, Reading, MA, (1995), Chap. 7.

\bibitem{WilliamsFallerHill:1971}
E. R. Williams, J. E. Faller, and H. A. Hill,
``New experimental test of Coulomb's law: 
A laboratory upper limit on the photon rest mass,''
Phys. Rev. Lett. {\bf 26}, 721--724 (1971).

\bibitem{GoldhaberNieto:1971}
A. S. Goldhaber and M. M. Nieto,
``Terrestrial and extra-terrestrial limits on the photon mass,''
Rev. Mod. Phys. {\bf 43}, 277--296 (1971).

\bibitem{Schwinger:1962}
J. Schwinger,
``Gauge Invariance and Mass II,''
Phys. Rev. {\bf 128}, 2425--2429 (1962).

\bibitem{Parker:1969}
L. Parker, Quantized fields and particle creation in expanding
universes. I,'' Phys. Rev. {\bf 183}, 1057--1068 (1969).

\bibitem{Gasiorowicz:1974}
S. Gasiorowicz, Quantum Physics (Wiley and Sons, Inc., 
New York, 1974), p. 361.

\bibitem{ProkopecTornkvistWoodard:2002}
T. Prokopec, O. T\"ornkvist and R. Woodard,
``Photon mass from inflation,''
Phys. Rev. Lett. {\bf 89}, 101301-1--101301-5 (2002)
or arXiv:astro-ph/0205331.

\bibitem{ProkopecTornkvistWoodard:2002b}
T. Prokopec, O. T\"ornkvist, and R. P. Woodard,
``One loop vacuum polarization in a locally de Sitter background,''
Ann. Phys. {\bf 303}, 251--274 (2003) or
arXiv:gr-qc/0205130.

\bibitem{DavisDimopoulosProkopecTornkvist:2000}
A.-C. Davis, K. Dimopoulos, T. Prokopec, and O. T\"orn\-kvist,
``Primordial spectrum of gauge fields from inflation,''
Phys. Lett. B {\bf 501}, 165--172 (2001) or
astro-ph/0007214.

\bibitem{TornkvistDavisDimopoulosProkopec:2000} O. T\"ornkvist,
A.-C. Davis, K. Dimopoulos, and T. \-Pro\-ko\-pec, ``Large scale
primordial magnetic fields from inflation and preheating,''
in {\sl Verbier 2000, Cosmology and particle physics} (CAPP2000), 
443--446, or astro-ph/0011278.

\bibitem{DimopoulosProkopecTornkvistDavis:2001} K. Dimopoulos, T.
Prokopec, O. T\"ornkvist, and A.-C. Davis, ``Natural magnetogenesis
from inflation,'' Phys. Rev. D {\bf 65}, 063505-1--063505-26
(2002) or astro-ph/0108093.

\bibitem{scalarvev} A. Vilenkin and L. H. Ford, ``Gravitational
effects upon cosmological phase transitions,'' Phys. Rev. D {\bf
26}, 1231--1241 (1982) ; A. D. Linde, ``Scalar field fluctuations
in expanding universe and the new inflationary universe scenario,''
Phys. Lett. B {\bf 116}, 335-339 (1982); A. A. Starobinsky,
``Dynamics of phase transitions in the new inflationary scenario and
generation of perturbations,'' Phys. Lett. B {\bf 117}, 175--178
(1982).

\bibitem{Kronberg:1994} P. P. Kronberg, ``Extragalactic magnetic
fields,'' Rept. Prog. Phys. {\bf 57}, 325--382 (1994).

\bibitem{GrassoRubinstein:2001} D. Grasso and H. R. Rubinstein,
``Magnetic fields in the early universe,'' Phys. Rept. {\bf 348},
163--266 (2001) or astro-ph/0009061.

\bibitem{dynamo} E. N. Parker, {\it Cosmic Magnetic Fields}
(Clarendon, Oxford, 1979); Ya. B. Zeldovich, A. A. Ruzmaikin and D.
D. Sokoloff, {\it Magnetic Fields in Astrophysics} (Gordon and
Breach, New York, 1983).

\bibitem{DavisLilleyTornkvist:1999} A.-C. Davis, M. Lilley, and O.
T\"ornkvist, ``Relaxing the bounds on primordial magnetic seed
fields,'' Phys. Rev. D {\bf 60}, 021301-1--021301-6 (1999) or
astro-ph/9904022.

\end{thebibliography}
\end{document}